\documentclass[dvips]{class}
\usepackage{supertabular,lscape,epsfig}
\usepackage{amssymb}
\usepackage{amsmath}
\usepackage{mathptmx}
\usepackage{booktabs}
\usepackage{rotating}
\usepackage{multirow}
\usepackage[a4paper, textwidth=13.7cm, textheight=21cm,
            hcentering=true, vcentering=true]{geometry}

\usepackage{txfonts}

\usepackage{xspace}
\usepackage{xcolor}
\usepackage{hyperref}

\newcommand{\new}[1]{{#1}}

\newcommand{\msun}{\mathrel{\mathrm{M}_\odot}}
\newcommand{\rsun}{\mathrel{\mathrm{R}_\odot}}

\newcommand{\myr}{\mathrel{\mathrm{Myr}}}
\newcommand{\gyr}{\mathrel{\mathrm{Gyr}}}
\newcommand{\kpc}{\mathrel{\mathrm{kpc}}}

\newcommand{\taueff}{\mathrel{\tau_\textrm{eff}}}

\newcommand{\msource}{\mathrel{\mathrm{m}_\mathrm{source}}}
\newcommand{\msourcesl}{\mathrel{\mathrm{m}_\mathrm{source,SL}}}
\newcommand{\musl}{\mathrel{\mu_\mathrm{sl}}}
\newcommand{\muslmax}{\mathrel{\mu_\mathrm{sl,max}}}
\newcommand{\porb}{\mathrel{\mathrm{P}_\mathrm{orb}}}
\newcommand{\texp}{\mathrel{\mathrm{t}_\mathrm{exp}}}
\newcommand{\tcadence}{\mathrel{\mathrm{t}_\mathrm{cadence}}}
\newcommand{\tsurvey}{\mathrel{\mathrm{t}_\mathrm{survey}}}
\newcommand{\pobs}{\mathrel{\mathrm{P}_\mathrm{obs}}}

\newcommand{\ML}{\mathrel{\mathrm{M}_\mathrm{lens}}}

\renewcommand{\RS}{\mathrel{\mathrm{R}_\mathrm{source}}}
\newcommand{\RL}{\mathrel{\mathrm{R}_\mathrm{lens}}}
\newcommand{\RE}{\mathrel{\mathrm{R}_\mathrm{E}}}
\newcommand{\rS}{\mathrel{\mathrm{r}_\mathrm{S}}}
\newcommand{\rL}{\mathrel{\mathrm{r}_\mathrm{L}}}

\newcommand{\sigmanoise}{\mathrel{\sigma_\mathrm{noise}}}

\newcommand{\npop}{\mathrel{\mathrm{N}_\mathrm{pop}}}
\newcommand{\fracb}{\mathrel{\mathrm{f}_\mathrm{bin}}}
\newcommand{\concpop}{\mathrel{\mathrm{R}_\mathrm{h,2}/\mathrm{R}_\mathrm{h,1}}}
\newcommand{\Erawa}{\mathrel{\mathbb{E}[\mathcal{N}(\musl > 1)]}}
\newcommand{\Erawb}{\mathrel{\mathbb{E}[\mathcal{N}(\musl > 2)]}}

\newcommand{\rh}{\mathrel{\mathrm{r}_\mathrm{h}}}
\newcommand{\frh}{\mathrel{f_{\mathrm{r}_\mathrm{h}}}}

\newcommand{\frc}{\mathrel{f_{\mathrm{r}_\mathrm{c}}}}
\newcommand{\Ecov}{\mathrel{\mathbb{E}[\mathcal{N}_\mathrm{cov}]}}
\newcommand{\Erec}{\mathrel{\mathbb{E}[\mathcal{N}_\mathrm{rec}]}}
\newcommand{\Etot}{\mathrel{\mathbb{E}[\mathcal{N}_\mathrm{tot}]}}

\newcommand{\Ks}{K$_{\rm S}$\xspace}

\newcommand{\ApJl}{ApJL}
\newcommand{\apss}{ApSS}
\newcommand{\pasp}{PASP}
\newcommand{\LRSP}{Living Rev. Solar Phys.}
\newcommand{\SPIE}{Proc.\ SPIE}

\newcommand{\figmdistribution}{1\xspace}
\newcommand{\figdmdistribution}{2\xspace}
\newcommand{\figtaueffdistribution}{3\xspace}
\newcommand{\figmfiltereddistributions}{4\xspace}
\newcommand{\figPobsdistribution}{5\xspace}
\newcommand{\figEtotdistributions}{6\xspace}
\newcommand{\figkdistribution}{7\xspace}
\newcommand{\figELTdistributions}{8\xspace}
\newcommand{\figAllELTdistributions}{9\xspace}

\newcommand{\tabresults}{1\xspace}
\newcommand{\tabeltaccessibility}{2\xspace}
\newcommand{\tabphotometricdetection}{3\xspace}
\newcommand{\tabsurveystrategy}{4\xspace}
\newcommand{\tabcomprehensivepredictions}{5\xspace}

\begin{document}

\begin{Titlepage}

\Title{Self-lensing binaries in globular clusters: theoretical characterization and predictions for ELT}

\Author{G.~~W~i~k~t~o~r~o~w~i~c~z$^1$,~~M.~~M~i~d~d~l~e~t~o~n$^2$,~~M.~~G~i~e~r~s~z$^1$,\\ ~~A. I~n~g~r~a~m$^3$,~~A.~~M~c~M~a~s~t~e~r$^2$,~~A.~~A~s~k~a~r$^1$,\\ and~~~L.~~H~e~l~l~s~t~r~\"o~m$^1$}{${}^1$ Nicolaus Copernicus Astronomical Center, Polish Academy of Sciences, Bartycka 18, 00-716 Warsaw, Poland; e-mail: gwiktoro@camk.edu.pl \\ ${}^2$ School of Physics \& Astronomy, University of Southampton, Southampton, Southampton SO17 1BJ, UK \\ ${}^3$ School of Mathematics, Statistics, and Physics, Newcastle University, Newcastle upon Tyne, NE1 7RU, UK }

\footnotesize{January, 2026}
\end{Titlepage}

\Abstract{Self-lensing (SL) represents a powerful technique for detecting compact objects in binary systems through gravitational microlensing effects, when a compact companion transits in front of its luminous partner. We present the first comprehensive study of SL probability within globular cluster (GC) environments, utilizing synthetic stellar populations from MOCCA simulations. We find that present-day GCs contain 1--50 SL sources with magnifications $\musl > 1+10^{-8}$, strongly dependent on initial binary fraction, with systems dominated by WD lenses paired with low-mass main-sequence companions. The predicted populations exhibit characteristic bimodal magnitude distributions with peaks at $m \approx 24$ and 32 mag at 10 kpc distance, and typical Einstein ring crossing times of $\taueff \sim 2$ hours. Our analysis incorporates finite-size lens effects for white dwarf (WD) lenses and the specific observational characteristics of the Extremely Large Telescope's (ELT) MICADO instrument. Multi-year monitoring campaigns with daily cadence provide order-of-magnitude improvements over \new{a random-epoch geometric baseline through increased temporal coverage and improved phase-folded photometric precision. Our results provide a framework for assessing the potential of ELT monitoring of Galactic GCs to detect otherwise hidden SL binaries.}}{binaries: general -- gravitational lensing: micro -- globular clusters: general -- methods: numerical -- stars: statistics}

\section{Introduction}

Self-lensing (SL) represents a powerful astronomical technique for detecting compact objects in binary systems, first proposed decades ago \new{(Maeder 1973, Gould 1995)} but gaining renewed attention due to advances in large-scale sky surveys and instrumental precision. The method exploits gravitational microlensing effects when a compact object transits in front of its luminous companion, producing characteristic photometric signatures that encode fundamental properties of the lensing system (Witt \& Mao 1994, Agol 2002).

The observational landscape for SL has evolved dramatically in recent years. Space-based missions such as Kepler and TESS have enabled the discovery of several SL systems featuring white dwarf (WD) lenses paired with main-sequence companions (Kruse \& Agol 2014, Kawahara \etal 2018, Masuda \& Hotokezaka 2019, Sorabella \etal 2024). The method was shown to be capable of testing stellar evolutionary physics (Wiktorowicz \etal 2025b) and avoid many of the selection biases inherent in traditional techniques for finding compact objects.

On the other hand, SL detection faces significant observational challenges. The method requires nearly edge-on binary configurations and wide orbits, historically considered prohibitively rare for systematic surveys. Additionally, finite-size effects introduce degeneracies between magnification and demagnification, particularly for WD lenses, which can obscure or completely hide the SL signal (Han 2016). Recent theoretical work has begun addressing these complications through improved modelling of finite-size lens effects \new{(Agol 2002, Sajadian 2023)}.

Globular clusters (GCs) represent a particularly promising but unexplored environment for SL searches. These stellar systems are expected to harbour significant populations of compact objects, including substantial numbers of double white dwarf binaries (DWDs), though such systems often remain individually undetectable due to their intrinsic faintness (\eg Hellstr{\"o}m \etal 2024). The high stellar densities and old ages characteristic of Galactic GCs suggest they may contain enhanced populations of evolved binary systems suitable for SL studies, yet no comprehensive investigation of SL in the GC environment has been undertaken.

The advent of the Extremely Large Telescope \new{(ELT; Davies \etal 2016)} and its first-generation instrument \new{MICADO (Davies \etal 2021)} provides the angular resolution needed to assess SL monitoring in crowded GC fields. With angular resolution approximately six times better than the James Webb Space Telescope \new{(Rigby \etal 2023)}, MICADO should permit photometric monitoring of individual stars within globular cluster cores, where source confusion has historically limited the detectability of faint variables and transients.

In this work, we present the first comprehensive study of SL potential within GC environments, extending previous methodologies (Wiktorowicz \etal 2021) to incorporate finite-size lens effects and the specific observational characteristics of the ELT. We utilize synthetic GC populations from the MOCCA Monte Carlo cluster evolution code \new{(Giersz 1998, Hypki \& Giersz 2013)} to model realistic stellar environments and predict the expected yield of detectable SL events (see Section~\ref{sec:MOCCA}). Our analysis focuses specifically on white dwarf lenses, which despite producing lower magnification amplitudes due to finite-size effects, offer the advantage of potential occultation signatures and represent the most abundant compact object population in old stellar systems.

This investigation represents a natural extension of recent population synthesis studies of Galactic SL systems, shifting focus from wide-field survey capabilities to the \new{high angular resolution} regime where ELT-class telescopes excel. While previous work emphasized the potential of all-sky surveys such as \new{LSST (Ivezi{\'c} \etal 2019) and ZTF (Bellm \etal 2019)} for discovering SL systems throughout the Galaxy \new{(Wiktorowicz \etal 2021, Wiktorowicz \etal 2025b)}, our approach targets the dense stellar environments where next-generation ground-based telescopes can achieve their greatest scientific impact.

\section{Methods}

\subsection{Self-Lensing}

All binary systems are positioned at the same distance as their host GC. Apparent magnitudes are calculated based on the cluster distance and the intrinsic luminosity of the stars. Orbital inclinations are sampled uniformly in $\cos i$ to ensure isotropic orientation distributions, consistent with the assumption of randomly oriented binary orbits (Heggie 1975). We account for uncertainties in positioning, inclination, and snapshot sampling through utilization of multiple snapshots of similar ages.

We define a "detectable" SL event as a binary system where at least one photometric observation can occur during the magnification event for a given instrument configuration. This criterion focuses on the geometric probability of temporal coincidence between the lensing event and observations, without requiring explicit consideration of photometric precision or the ability to distinguish SL from other variable phenomena.

Following the approach of (Wiktorowicz \etal 2025b), this geometric detectability serves as an upper limit to the true detection rate. More conservative estimates that account for photometric noise and stellar crowding are presented in Section~\ref{sec:conservativepredictions}.

The application of SL techniques to GC environments introduces several complications absent in Galactic field studies. Unlike neutron star or black hole lenses where point-source approximations are typically valid, WD lenses (which dominate in old stellar populations) in typical binary configurations often have physical radii comparable to their Einstein radii, requiring detailed treatment of finite-size effects.

SL involving finite-size lenses represents a phenomenon where both gravitational lensing and occultation effects become significant (Agol 2002). The transition between pure lensing and occultation regimes occurs when the radius of the lens star ($\RL$) becomes comparable to the Einstein radius ($\RE$).

Following (Agol 2002), gravitational lensing by a point mass produces two images of a distant source: one interior and one exterior to the Einstein radius in the lens plane, defined as:

\begin{equation}
    \RE = \sqrt{\frac{4G\ML a \sin i}{c^2}}
\end{equation}

\noindent where $\ML$ is the lens mass, $a$ is the separation, and $i$ is the inclination, with $G$ and $c$ denoting the gravitational constant and speed of light, respectively.

The ratio $\rL = \RL/\RE$ determines the lensing regime:

\begin{itemize}
    \item $\rL < 1$ (Small lens regime): The inner image may be partially occulted when the source approaches the lens, while the outer image remains unocculted. The inner image appears (unoccults) as the source-lens separation increases.
    \item $\rL > 1$ (Large lens regime): The inner image is always fully occulted, and the outer image may be occulted as the source approaches the lens. This produces characteristically flatter lightcurve profiles near maximum magnification.
\end{itemize}

For extended sources with normalized radius $\rS = \RS / \RE$, we implement the analytical prescriptions of Agol (2002). The magnification is computed as the ratio of the unocculted area of the lensed images to the area of the unlensed source, since surface brightness is conserved during lensing.

The total magnification for a uniform source is given by:
\begin{equation}
    \musl = \mu_\text{sl,+} + \mu_\text{sl,-}
\end{equation}

\noindent where $\mu_\text{sl,+}$ and $\mu_\text{sl,-}$ represent the magnifications of the outer and inner images, respectively\footnote{We note that nomenclature conventions for SL magnification are not yet standardized in the literature. Some authors use $A$ following microlensing conventions (\eg Han 2016), while others adopt $s_\mathrm{sl}$ (\eg Masuda \& Hotokezaka 2019). In this work, we follow the $\mu$ notation established in the foundational SL literature, \eg Witt \& Mao (1994), Agol (2002), Agol (2003), with the subscript 'sl' explicitly denoting SL to distinguish it from other physical quantities. This notation maintains consistency with our previous studies (Wiktorowicz \etal 2021, Wiktorowicz \etal 2025b).}. The expressions for these magnifications depend on the specific combination of $\rL$, $\rS$, and source position $\zeta_0:=b/\RE$ (i.e. the impact factor $b$ in units of Einstein radius $\RE$), as detailed in Tables 1 and 2 of Agol 2002.

In cases where occultation effects are negligible ($\rL \ll 1$), this implementation reduces to the standard point-like lens prescriptions of Witt \& Mao (1994), ensuring consistency with previous studies of NS and BH lenses (\eg Wiktorowicz \etal 2021, Wiktorowicz \etal 2025b).

The finite-size lens formalism is particularly relevant for WD lenses, whose physical radii are often comparable to their Einstein radii in typical binaries. The "demagnification" caused by occultation of the inner image (and sometimes the outer image) can lead to significant reduction in total magnification, and in extreme cases, can reduce the apparent source luminosity below its baseline level.

Following the classification scheme of Sajadian (2025), we distinguish between different SL extremes for WD lenses:
\begin{itemize}
    \item Strong Lensing: Occurs in systems with two massive WDs in wide orbits, maximizing $\RE$ while minimizing the relative lens size, thereby reducing occultation effects.
    \item Deep Eclipses: Dominate in systems with low-mass WD lenses and massive sources in close orbits, causing deep or complete eclipses that overwhelm the lensing signal.
\end{itemize}

We assume all lens stars to be extended objects and apply the finite-size lens prescriptions throughout our analysis. For computational efficiency, we adopt a uniform source brightness approximation, neglecting limb-darkening effects (\eg Claret 2000, Agol 2002, Lee \etal 2009, Sajadian \& Afshordi 2024). This approximation is justified for survey-level detectability studies, though more detailed modelling would require numerical integration of limb-darkened profiles.

We estimate the lensing duration, building on the formalism of Masuda \& Hotokezaka (2019) and Wiktorowicz \etal (2021), and define the effective crossing time as
\begin{equation}
    \taueff = \frac{\porb}{\pi a}\sqrt{(\RE + \RS)^2 - b^2},
\end{equation}
\noindent where $\porb$ is the orbital period and $b\;[\rsun]$ is the impact parameter, that is, the distance between the source and lens centres in the lens plane. This expression uses the circular-orbit speed at the semimajor axis and therefore approximates eccentric systems.

\subsection{MOCCA simulations}\label{sec:MOCCA}

The MOCCA (MOnte Carlo Cluster evolution Code) is an advanced numerical tool designed to simulate the complete stellar and dynamical evolution of realistic star clusters, including GCs, over cosmological timescales (Giersz 1998, Hypki \& Giersz 2013, Giersz \etal 2013). The code combines Monte Carlo methods for stellar dynamics (H{\'e}non 1971, Stodolkiewicz 1986) with detailed stellar and binary evolution algorithms from the Binary Star Evolution code (BSE; Hurley \etal 2000, Hurley \etal 2002; with later updates, \eg  Belloni \etal 2018, Kamlah \etal 2022) and low-N scattering code FEWBODY (Fregeau \etal 2004) to follow close dynamical interactions. This hybrid approach provides an optimal balance between computational efficiency and physical accuracy, making MOCCA particularly well-suited for studying complex astrophysical phenomena within dense stellar environments.

Recent improvements to the MOCCA code have significantly enhanced its capabilities, including sophisticated treatment of super-Eddington accretion, improved core radii calculations for Roche-lobe overflow, detailed evolutionary tracking, and the capability to handle multiple stellar populations with distinct chemical compositions, ages, and spatial distributions (Hypki \etal 2022, Hypki \etal 2025, Hellstr{\"o}m \etal 2025, Giersz \etal 2025). These capabilities enable MOCCA to provide valuable insights into the formation scenarios and long-term evolution of GCs, including the mechanisms that shape the observed spatial distributions and kinematic properties of their stellar populations.

For this study, we use the simulation dataset in Wiktorowicz \etal (2025a), which includes the most important updates to the MOCCA code and represents a comprehensive range of GC properties representative of Galactic systems. The dataset comprises 20 simulations with dynamical interactions enabled, spanning various initial conditions including different numbers of stellar populations (1 or 2), binary fractions (10\% or 95\%), and tidal filling configurations (tidally filling or non-tidally filling).

For the 95\% binary fraction cases, we employ the Kroupa (1995) initial binary parameter distribution (modified by Belloni \etal 2017) for stars below $5\msun$, which includes wide/long period binaries that are particularly relevant for this study. For the 10\% binary fraction cases, we use a uniform distribution in logarithm for the semi-major axis, ranging from the minimum distance (sum of stellar radii) up to $100$ AU. Stars above $5\msun$ in both cases follow the Sana \etal (2012) distributions, with approximately flat mass ratio distributions.

All clusters were initialized with Kroupa initial mass functions (mass range 0.08--$150\msun$; Kroupa 1995), King density profiles (King 1962) with concentration parameters $W_0 = 3.0$ and $W_0 = 7.0$ for first and second populations respectively, and metallicity characteristic of Galactic globular clusters ($Z = 0.001$). 
For complete details of the simulation parameters and initial conditions, we refer the reader to Wiktorowicz \etal (2025a).

\new{We emphasize that MOCCA models binary encounters throughout the evolution, with strong few-body interactions handled using the integrated FEWBODY toolkit (Fregeau \etal 2004). The present-day binary populations therefore include the accumulated effects of dynamical hardening, softening, exchanges, and disruption over the full $12\gyr$ cluster lifetime (Hypki \& Giersz 2013, Giersz \etal 2013).}

In order to obtain present-day stellar populations (their masses, radii, bolometric luminosities, etc.), we extract cluster properties at an age of $12\gyr$. To improve statistical robustness for our analysis, we average results from snapshots between 11--$13\gyr$. This approach effectively provides 5 snapshots per simulation, yielding a total of 100 cluster realizations. This averaging is justified since GCs evolve relatively slowly at such advanced ages, with minimal structural changes occurring over this $2\gyr$ interval.

All results presented in this work (including the numbers of SL binaries in the tables) correspond to individual simulated clusters and are not scaled to represent the total observable population of Galactic GCs.

\subsection{The Extremely Large Telescope}\label{sec:ELT}

The Extremely Large Telescope (ELT) is a 39-meter ground-based telescope designed to achieve diffraction-limited angular resolution of $\sim7$ mas at $1.2 \mu$m (Leschinski \& Alves 2020). This resolution is approximately six times better than that of JWST and provides an effective sensitivity gain of $\sim3$ magnitudes in crowded stellar fields (Padovani \& Cirasuolo 2023, Davies \etal 2016).

MICADO, the first-generation near-infrared imager covering 0.8--2.4 $\mu$m (Davies \etal 2021), features two complementary modes: a wide-field imager (50.5" $\times$ 50.5" FoV, 4 mas pixels) and a high-resolution imager (19" $\times$ 19" FoV, 1.5 mas pixels). For this study, we focus on the high-resolution mode, which ensures complete sampling of the diffraction-limited PSF and optimal photometric precision in dense stellar environments.

MICADO achieves $5\sigma$ point-source sensitivities of J$_{\text{AB}} \sim 28.6$, H$_{\text{AB}} \sim 29.5$, and K$_{\text{AB}} \sim 29.1$ mag for 5-hour integrations, with practical detection limits of \Ks = 28 mag in crowded fields (Davies \etal 2021, Leschinski \& Alves 2020). The instrument supports high-frequency sampling (even sub-second exposures; Davies \etal 2016), crucial for the rapid variability characteristic of SL light curves.

\new{An important observational constraint is MICADO's limited field of view. In high-resolution mode (19'' $\times$ 19'' FoV), a single pointing covers only a fraction of a GC's angular extent, particularly for nearby ($D\lesssim10\kpc$) clusters. A core radius of $r_c = 0.5$--$1$ pc (the median for Galactic GCs is $r_c \approx 0.8$ pc; Baumgardt \& Hilker 2018) subtends an angular radius of $21''$--$41''$ at 5 kpc, so a central pointing covers $\sim7$--$27\%$ of the projected core area. At 10 kpc, one pointing covers most, but not all, of a compact core with $r_c\approx0.5$ pc. To quantify central concentration independently of pointing geometry, Table~\tabresults reports the $\Erawa$-weighted projected fraction of SL sources inside each model's MOCCA core radius, $\frc$; it is typically around $20\%$ and reaches $45\%$. Because $\frc$ is not the fraction inside the square MICADO FoV, we do not use it to rescale the predicted rates. The rates in Section~\ref{sec:eltpredictions} remain whole-cluster values and are therefore upper limits for a single pointing.}

\section{Results}

\subsection{Expected populations and magnifications}

Below we present results for all SL sources in our sample. We limit our analysis to systems in which the projected source intersects the Einstein ring, providing an effective limit on magnification of $\musl - 1 > 10^{-8}$. Additionally, we exclude systems where occultation reduces the luminosity more than SL magnification enhances it, resulting in net dimming or no net brightening ($\musl \leq 1$).

\begin{sidewaystable*}
\centering
\caption{Simulation parameters and self-lensing predictions for various globular cluster evolution models}
\begin{tabular}{lrlrlllllll}
\toprule
Label & $\npop$ & $\concpop$ & $\fracb$ & TF & NF & $\Erawa$ & $\Erawb$ & $\muslmax$ & $\frh$ & $\frc$ \\
\midrule
npop1-fb10-TF & 1 &  & 10 & T &  &  &  &  &  & \\
npop1-fb10-nTF & 1 &  & 10 &  &  & 4.0(2) & 0.012(1) & 26(12) & 0.91(1) & 0.27(24) \\
npop1-fb95-TF & 1 &  & 95 & T &  &  &  &  &  & \\
npop1-fb95-nTF & 1 &  & 95 &  &  & 28(1) & 0.036(6) & 28(12) & 0.89 & 0.23(12) \\
npop2-cpop05-fb10-TF & 2 & 0.05 & 10 & T &  & 1.4(2) & 0.0100(17) & 43(16) & 0.77(4) & 0.34(23) \\
npop2-cpop05-fb10-TF-NF & 2 & 0.05 & 10 & T & T & 7.3(5) & 0.045(3) & 101(22) & 0.61(3) & 0.01(1) \\
npop2-cpop05-fb10-nTF & 2 & 0.05 & 10 &  &  & 2.8(2) & 0.0066(23) & 17(8) & 0.80(5) & 0.19(7) \\
npop2-cpop05-fb10-nTF-NF & 2 & 0.05 & 10 &  & T & 1.3(1) & 0.0048(19) & 16(5) & 0.87(4) & 0.21(5) \\
npop2-cpop05-fb95-TF & 2 & 0.05 & 95 & T &  & 9.5(3) & 0.014(1) & 36(10) & 0.81(2) & 0.45(14) \\
npop2-cpop05-fb95-TF-NF & 2 & 0.05 & 95 & T & T & 40(1) & 0.071(2) & 177(76) & 0.55(1) & 0.02(1) \\
npop2-cpop05-fb95-nTF & 2 & 0.05 & 95 &  &  & 16(1) & 0.015(6) & 17(6) & 0.77(2) & 0.20(5) \\
npop2-cpop05-fb95-nTF-NF & 2 & 0.05 & 95 &  & T & 9.9(5) & 0.0081(38) & 9.7(16) & 0.81(1) & 0.18(3) \\
npop2-cpop2-fb10-TF & 2 & 0.2 & 10 & T &  &  &  &  &  & \\
npop2-cpop2-fb10-TF-NF & 2 & 0.2 & 10 & T & T & 8.0(2) & 0.057(3) & 120(44) & 0.61(3) & 0.01(1) \\
npop2-cpop2-fb10-nTF & 2 & 0.2 & 10 &  &  & 3.6(4) & 0.0059(18) & 28(11) & 0.89(3) & 0.23(10) \\
npop2-cpop2-fb10-nTF-NF & 2 & 0.2 & 10 &  & T & 0.76(23) & 0.0093(28) & 7.9(46) & 0.82(9) & 0.13(7) \\
npop2-cpop2-fb95-TF & 2 & 0.2 & 95 & T &  &  &  &  &  & \\
npop2-cpop2-fb95-TF-NF & 2 & 0.2 & 95 & T & T & 52(1) & 0.085(2) & 176(105) & 0.59(1) & 0.02(2) \\
npop2-cpop2-fb95-nTF & 2 & 0.2 & 95 &  &  & 20(1) & 0.017(5) & 14(2) & 0.84(1) & 0.18(7) \\
npop2-cpop2-fb95-nTF-NF & 2 & 0.2 & 95 &  & T & 8.6(10) & 0.097(9) & 11(1) & 0.93(1) & 0.25(5) \\
\bottomrule
\end{tabular}

{Each label specifies: number of stellar populations ($\npop$), concentration parameter ($\concpop$), binary fraction ($\fracb$), initial tidal filling (TF), and inclusion of new features (NF). Results include expected numbers of self-lensing sources with magnifications $\musl > 1$ ($\Erawa$) and $\musl > 2$ ($\Erawb$), the highest magnification ($\muslmax$), and projected fractions of self-lensing sources inside the half-mass radius ($\frh$) and the model-specific MOCCA core radius ($\frc$). Empty cells indicate cluster dissolution before $11\gyr$. Sixteen out of 20 simulated clusters survive until that time. 1-$\sigma$ uncertainties are given in parentheses as errors in the last significant digit(s). Detailed simulation methodology is presented in Appendix A of Wiktorowicz \etal 2025a.}
\end{sidewaystable*}

Table~\tabresults summarizes the results of our simulations, detailed in Wiktorowicz \etal (2025a), with key parameters provided for reference. The expected number of SL sources with magnification $\musl > 1$ ($\Erawa$) ranges from approximately 1 to over 50 in a simulated cluster, whose parameters are in the observed range for Milky-Way GCs. Results are strongly influenced by the initial binary fraction ($\fracb$); simulations with $\fracb = 95\%$ consistently yield higher $\Erawa$ values than those with $\fracb = 10\%$. For example, in the npop2-cpop05 configuration, $\Erawa$ increases from $2.79\pm0.17$ to $15.99\pm0.93$ in non-tidally filling (nTF) systems, and from $1.36\pm0.19$ to $9.54\pm0.31$ in tidally filling (TF) systems as $\fracb$ rises from $10\%$ to $95\%$.

The inclusion of "new features" (NF, \eg delayed, second population formation; Giersz \etal 2025; for details see Wiktorowicz \etal 2025a) has contrasting effects depending on tidal filling status. In TF systems, NF increases $\Erawa$ by factors of 4--5 and boosts maximum magnification ($\muslmax$), often exceeding 100. In contrast, for nTF systems, NF typically reduces $\Erawa$ and results in lower $\muslmax$ values, likely due to increased disruption of wide binaries in tidally underfilled clusters (Heggie \& Hut 2003).

The initial GC structure also influences SL numbers. Configurations with more extended second stellar populations ($\concpop = 0.2$) produce higher $\Erawa$ values than more concentrated ones ($\concpop = 0.05$), particularly in TF-NF simulations, where $\Erawa$ increases from $40.09\pm1.15$ to $52.13\pm0.92$ for $\fracb = 95\%$. This suggests that less centrally concentrated populations reduce dynamical interactions in the core, preserving more SL binaries.

The expected number of systems with $\musl > 2$ ($\Erawb$) remains low ($\lesssim 0.1$), indicating that strongly magnified SL events are rare and require coordinated observations of multiple GCs for statistically significant detection rates. However, SL sources can achieve extreme magnifications, with $\muslmax$ exceeding 100 in some TF-NF simulations. These events typically involve wide orbital separations (100--1,000 AU) and nearly edge-on inclinations ($i \approx \pi/2$). They reside mostly in the outskirts of GCs as such conditions are dynamically unstable in dense GC cores due to frequent encounters (Heggie \& Hut 2003). The highest $\muslmax$ values occur in TF-NF systems, suggesting that tidally filling clusters with delayed second population formation, create favourable conditions for extreme magnification events, possibly because of strongly reduced mass loss and increased interaction rate compared to the TF models (Giersz \etal 2025).

SL sources are concentrated toward the GC center, with $55$--$93\%$ of systems residing within the half-mass radius ($\rh$). These sources typically involve lenses descended from massive stars, which migrate to the core via mass segregation (Spitzer 1969, Baumgardt \& Makino 2003). However, the dense core environment may suppress SL source populations by disrupting wide binaries through dynamical interactions, which either harden tight binaries or dissolve wider ones. Dynamical formation of SL binaries via three-body encounters may counteract this, particularly in nTF systems (Fregeau \etal 2004), although such systems would likely be extremely tight and therefore produce sources with only minimal magnifications.

Simulations with missing predictions in Table~\tabresults correspond to clusters that dissolved before reaching $11\gyr$, highlighting the critical role of long-term dynamical stability in sustaining significant SL populations in present-day GCs. Consequently, only the 16 surviving simulations are included in the further analysis.

\subsection{Parameter distributions of self-lensing sources}

\begin{figure}
    \centering
    \includegraphics[width=\linewidth]{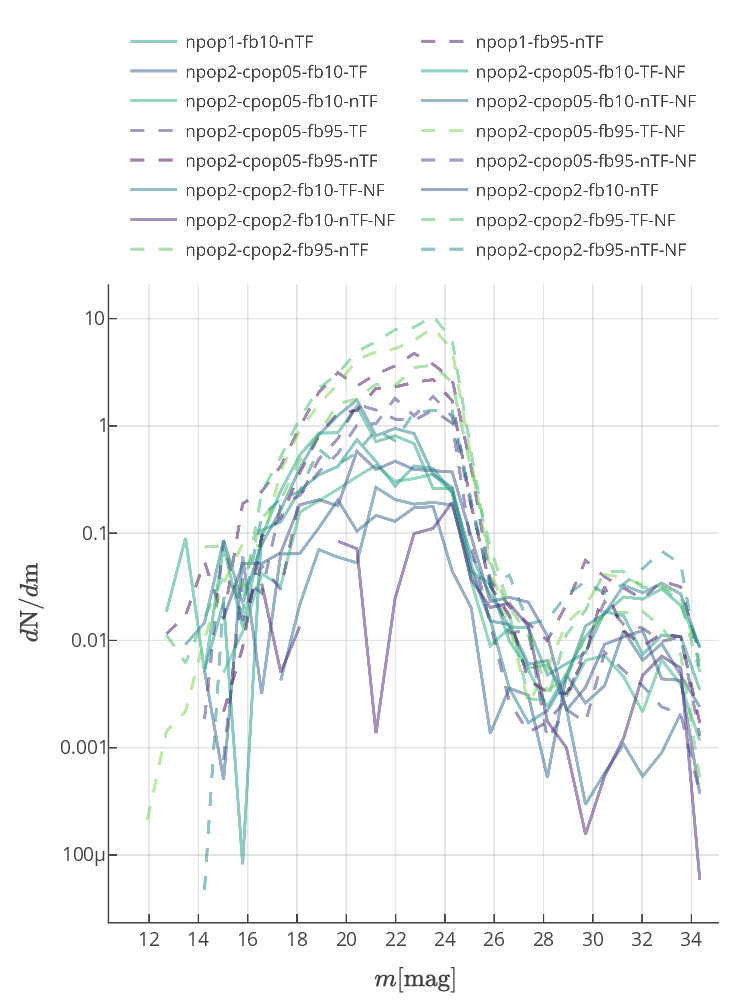}
    \FigCap{Apparent magnitude distribution of magnified self-lensing systems across 16 simulations, calculated for a reference distance of 10 kpc without extinction correction.}
\end{figure}

Figure~\figmdistribution presents the magnitude distributions for our 16 simulation configurations at the peak of SL magnification, calculated at a fiducial distance of $10\kpc$ without interstellar extinction. This distance is close to the median distance of Galactic globular clusters\footnote{https://people.smp.uq.edu.au/HolgerBaumgardt/globular/parameter.html} ($\sim 9.14$ kpc), \new{whose distances span approximately 1.8--150 kpc (Baumgardt \& Hilker 2018). Throughout this work, we adopt fiducial distances of 1, 10, and 100 kpc for computational convenience. The nearest known GC is M4 at $\sim 1.8$ kpc (Harris 1996), so the 1 kpc case is an illustrative lower bound rather than a prediction for a specific cluster. Apparent magnitudes can be rescaled using the distance modulus, whereas cluster-specific expectations additionally depend on extinction, structure, and crowding}.

The distributions exhibit a pronounced bimodal structure with a primary peak at $m \approx 24$ mag and a secondary peak at $m \approx 32$ mag. This bimodality appears consistently across all simulation configurations, suggesting a fundamental property of SL populations in GCs. The primary peak at 24 mag corresponds to relatively bright sources accessible to current ground-based surveys and space telescopes, while the fainter peak at 32 mag represents systems that would require next-generation facilities like the Extremely Large Telescope.

The relative prominence of these peaks varies systematically with initial binary fraction. Systems with higher initial binary fractions ($\fracb = 95\%$) show enhanced populations of high-SL sources compared to low binary fraction systems ($\fracb = 10\%$). In clusters with initially low binary fractions, three-body encounters in the dense core environment can dynamically assemble SL configurations involving more massive stellar remnants, which have segregated to the cluster centre; binaries created through this process are typically tight and consequently show low SL magnifications.

An important factor influencing magnification amplitudes is the evolutionary history of binary systems. "Pristine" systems that have retained their original companions since cluster formation exhibit systematically higher magnified luminosities than dynamically "altered" systems that have undergone exchanges, dissolutions or three-body formation processes. Our results show that only $2.3\pm2.7\%$ of "pristine" SL binaries have $\msourcesl>26$ while it grows to $4.6\pm2.2 \%$ of "altered" ones. This distinction partly explains the enhanced SL numbers observed in GCs with higher initial binary fraction.

\begin{figure}
    \centering
    \includegraphics[width=\linewidth]{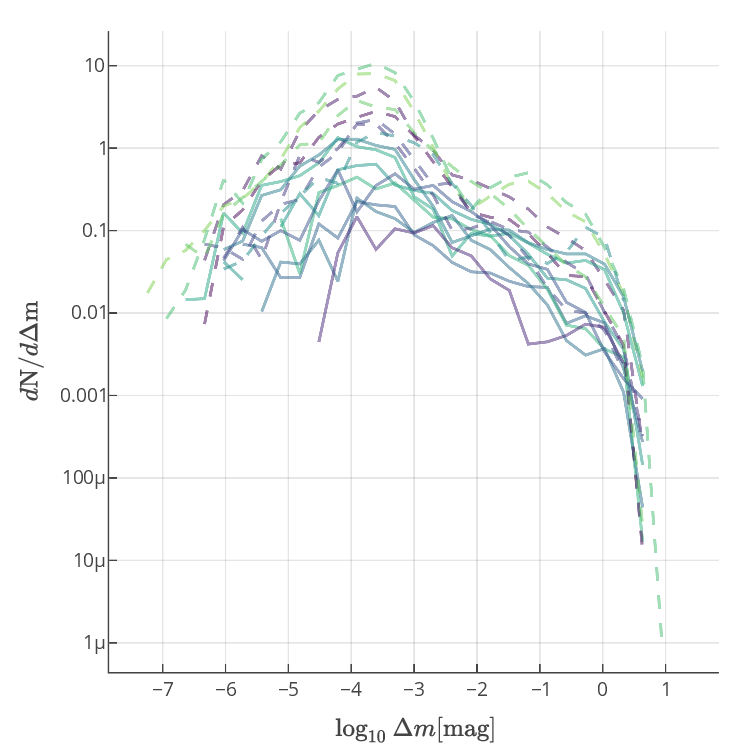}
    \FigCap{Distribution of magnifications for self-lensing systems across 16 simulations, expressed as magnitude differences $\Delta m = \msource - \msourcesl$ where positive values indicate brightening due to lensing. Labels as in Figure~\figmdistribution.}
\end{figure}

Figure \figdmdistribution presents the distribution of magnitude differences $\Delta m =  \msource - \msourcesl = 2.5\log_{10}\musl$. All simulations exhibit a pronounced peak at $\Delta m \approx 10^{-4}$ mag, indicating that the majority of SL sources experience only minimal magnification. These low-amplitude events likely fall below the detection threshold of current photometric surveys due to instrumental noise and atmospheric variability. The peak appears in all simulation configurations examined, although its physical origin is not isolated by the present analysis.

For magnitude changes $\Delta m > 10^{-3}$ mag, GCs with initially high binary fractions ($\fracb = 95\%$) systematically produce higher relative numbers of detectable SL sources compared to clusters with initially low binary fraction ($\fracb = 10\%$). This trend indicates that the adopted initial binary population affects the high-amplitude tail. In clusters with low primordial binary fractions, three-body encounters in the dense core environment preferentially assemble tight systems involving massive stellar remnants that have undergone mass segregation. These dynamically formed binaries rarely possess the large orbital separations necessary for significant lensing magnifications.

Systems with $\Delta m > 1$ mag are rare in the simulated population. The largest simulated magnitude changes reach approximately 5 mag.

Figure~\figkdistribution presents the distributions of lens and source stellar evolutionary types across our simulation suite. The results reveal systematic patterns that reflect both the stellar evolution timescales and dynamical processes operating within these dense stellar environments.

WD lenses dominate the SL population, accounting for the majority of predicted systems across all configurations. Among WD subtypes, carbon-oxygen white dwarfs are most prevalent, reflecting the typical evolutionary endpoints of intermediate-mass stars that constitute the bulk of GC stellar populations that managed to evolve off the main sequence within the assumed cluster age of $12\gyr$. Such stars segregate rather slowly. The abundance of WD lenses is expected given the evolved state of stellar systems like GCs, where stars with initial masses $\lesssim 8\msun$ (depending on metallicity and interactions) have had sufficient time to complete their evolution over the cluster's $\sim 12\gyr$ age, though the exact distribution depends on the presence of BH subsystem and intermediate-mass BH effects on stellar evolution and dynamics.

NSs represent the second most common lens type, consistently outnumbering BH lenses across all simulations. This trend aligns with expectations from both Galactic field populations (Wiktorowicz \etal 2021, Wiktorowicz \etal 2025b) and theoretical predictions for stellar remnant formation in GCs. The NS-to-BH ratio reflects the steep initial mass function in GCs with more NS-progenitors from a narrow mass range ($8{-}21\msun$) than the wider range ($\gtrsim 21\msun$) required for BH production. Many massive stars that would otherwise form NSs or BHs may have been ejected from GCs through dynamical interactions or natal kicks during their formation. This process is more efficient for lighter NSs which typically obtain stronger natal kicks.

The source stellar type distribution is strongly dominated by MS stars, particularly low-mass systems with $M < 0.7\msun$. These M/K-dwarf sources appear consistently across all lens categories -- WDs, NSs, and BHs -- reflecting their overwhelming  dominance in GC populations and their exceptionally long MS lifetimes ($\gtrsim 10\gyr$). The prevalence of low-mass MS sources also ensures high intrinsic detectability for SL events, as these stars provide stable, long-duration signals suitable for sustained photometric monitoring of periodic SL flares.

Evolved stellar sources show a distinct pattern reflecting evolutionary timescales and stellar structure considerations. Hertzsprung gap and red giant stars appear frequently as sources, benefiting from both their expanded photospheres (which increase lensing cross-sections) and their significant representation in present-day GC populations. In contrast, asymptotic giant branch stars and helium stars are notably under-represented despite their large radii. This scarcity reflects their brief evolutionary phases ($\lesssim 1\myr$ for thermal pulse episodes) and the low probability of capturing these transient evolutionary states in our $12\gyr$ snapshot simulations.

A significant fraction of sources are themselves WDs, particularly in double WD configurations (see also Hellstr{\"o}m \etal 2024). These systems represent an important subset of SL populations where both components can act as lens or source (Sajadian 2025). 

\begin{figure}
    \centering
    \includegraphics[width=\linewidth]{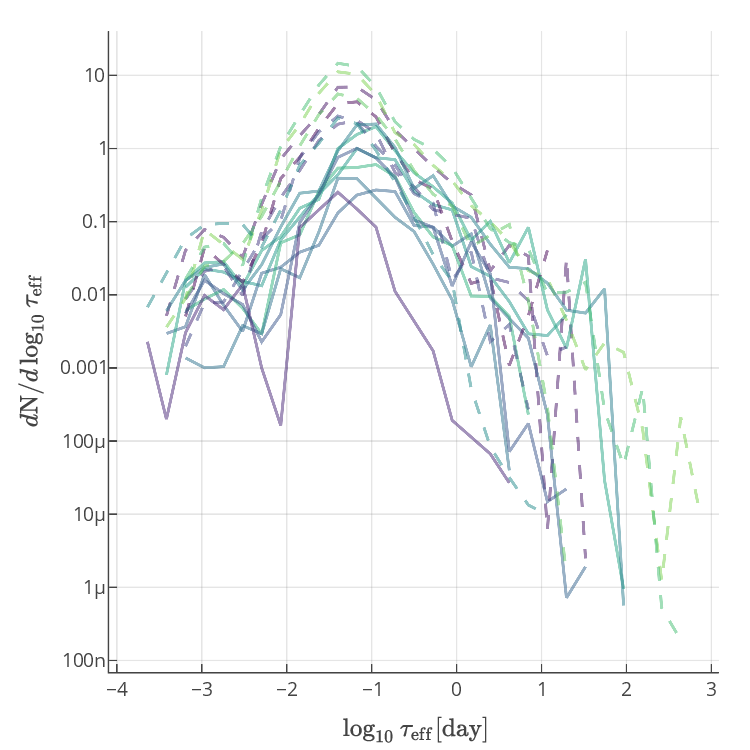}
    \FigCap{Distribution of Einstein ring crossing times ($\taueff$) for magnified self-lensing systems across all simulation configurations. Labels as in Figure~\figmdistribution.}
\end{figure}

The temporal characteristics of SL events are important for assessing cadence requirements. Figure~\figtaueffdistribution presents the distribution of effective Einstein ring crossing times ($\taueff$; see equation (5) of Wiktorowicz \etal 2021) for magnified SL systems across 16 simulated GCs. The $\taueff$ parameter represents the full event-duration chord crossing through the effective lensing region.

The distributions exhibit consistency across different simulation parameters, indicating that the temporal properties of SL events are primarily governed by the fundamental physics rather than the specific evolutionary history of individual clusters. The distribution is characterized by a sharp peak at $\taueff \sim 2$ hours, with approximately $50\%$ of events having durations between 30 minutes and 8 hours.

This short typical timescale imposes stringent constraints on observational cadence requirements. The majority ($\sim 80\%$) of SL events have durations shorter than 1 day, demanding high-frequency monitoring programs with cadences of hours or less to achieve reasonable detection efficiencies. This requirement is particularly challenging for ground-based surveys, which are often limited by weather conditions, telescope scheduling constraints, and the need to monitor multiple targets simultaneously.

The distribution exhibits an extended tail toward longer durations, with some events reaching $\taueff \gtrsim 100$ days. These long-duration systems are rare in the simulated population, but their extended durations could facilitate detailed follow-up.

\subsection{Predictions for ELT}\label{sec:eltpredictions}

We use the anticipated characteristics of ELT/MICADO to estimate the detectability of SL sources in globular clusters. The calculations provide idealized benchmarks based on current instrument specifications (Section~\ref{sec:ELT}) and can be refined as those specifications develop.

\subsubsection{Filter Selection and Magnitude Limits}

We evaluate ELT/MICADO performance in the J and \Ks bands. Stellar magnitudes were calculated assuming black-body radiation with effective temperatures derived from stellar evolution models, and synthetic photometry was performed using filter transmission curves from the ScopeSim package (Leschinski \etal 2020)\footnote{Filter profiles available at: \url{https://github.com/AstarVienna/irdb/tree/master/MICADO/filters}}. We note that these represent manufacturer estimates rather than final instrument specifications.

For SL detections we adopt a 3-$\sigma$ significance criterion (SNR=3). This corresponds to effective magnitude limits of J = 29.15 mag and \Ks = 28.55 mag, calculated using the relation $\Delta m = -2.5 \log_{10}(\text{SNR}/\text{SNR}_{\rm ref})\approx0.55$ [mag], where SNR$_{\rm ref}=5$. \new{These limits correspond to 5-hour integrations and characterize the deep-imaging sensitivity. For the adopted 1-minute benchmark visits, background-limited $\sqrt{t}$ scaling gives a limit shallower by $2.5\log_{10}\sqrt{5\,{\rm h}/1\,{\rm min}} \approx 3.1$ mag, i.e. \Ks $=25.46$ mag. This is a sensitivity benchmark rather than an operational exposure-time or scheduling assessment.}

\begin{figure}
\centering
\includegraphics[width=\linewidth]{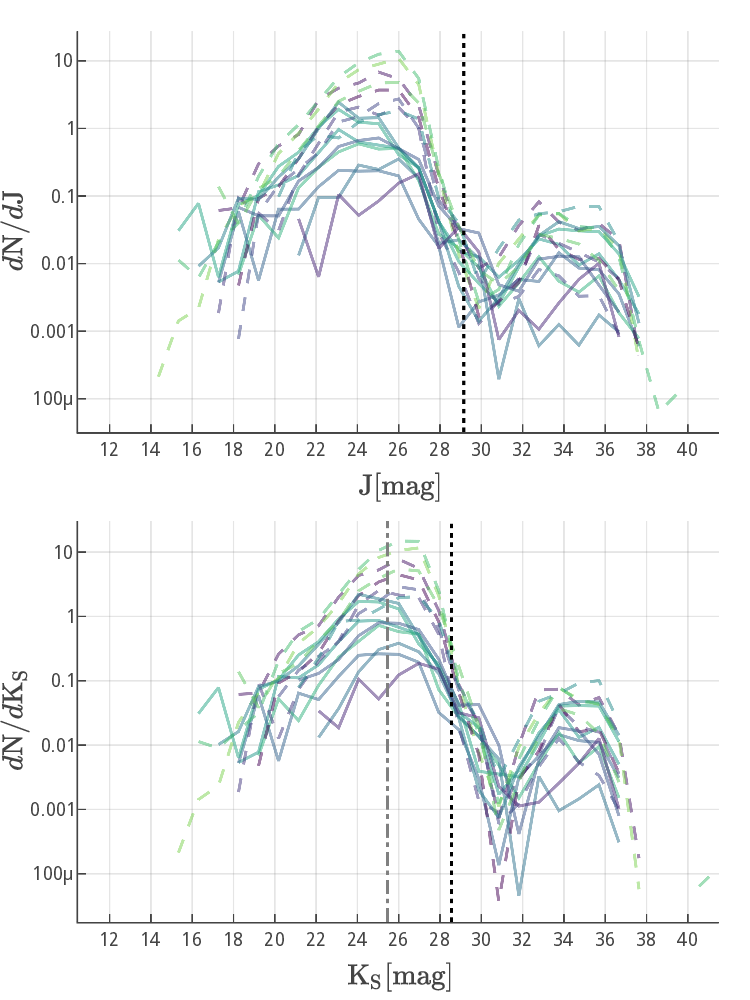}
\FigCap{Apparent magnitude distributions of self-lensing sources in J-band (top) and \Ks-band (bottom) filters for ELT/MICADO observations, calculated at a reference distance of 10 kpc. Vertical dashed lines show the 5-hour, 3-$\sigma$ limits of J = 29.15 mag and \Ks = 28.55 mag; \new{the additional dash-dotted line in the \Ks panel marks the adopted 1-minute survey-mode limit, \Ks $=25.46$ mag}. Labels as in Figure~\figmdistribution.}
\end{figure}

Figure~\figmfiltereddistributions presents the magnitude distributions for SL sources in both filters, calculated for the distance of 10 kpc without interstellar extinction corrections. The distributions maintain the characteristic bimodal structure observed in the broad-band case (Figure~\figmdistribution), with primary peaks at J $\approx$ 25 mag and \Ks $\approx$ 24 mag, and secondary \new{peaks around 32--34 mag, below the detection threshold.}

The bimodal magnitude distribution in Figure~\figmfiltereddistributions reflects the evolutionary demographics of 12 Gyr-old GC populations. The primary peak at $m\approx24$ mag arises from WD+low-mass MS binaries, where an evolved carbon-oxygen WD ($\sim0.6$--$0.8\msun$) lens has a faint K/M-dwarf companion ($\sim0.3$--$0.5\msun$). This configuration dominates numerically because low-mass stars greatly outnumber evolved remnants due to the assumed IMF, and WDs are the most abundant compact objects in old clusters where NS/BH populations have been depleted through natal kicks and dynamical ejections.

The secondary peaks at $m\approx32$--34 mag represent double WD systems, where both components are intrinsically faint, but maintain significant SL magnifications. These systems are rare and systematically fainter than WD+MS configurations by $\sim8$ magnitudes, creating the distinct secondary feature.

Distance scaling strongly affects ELT accessibility of GC populations. \new{For the adopted 1-minute limit, the primary magnitude peak satisfies the single-visit magnitude criterion out to approximately 20 kpc. For more distant Galactic clusters, only the brightest tail remains above this limit.}

The brightest SL sources may approach the saturation limit of \Ks $\sim 14.8$ mag for 2.6-second exposures (Leschinski \& Alves 2020). However, our predicted source populations are dominated by intrinsically faint objects (old MS stars and WDs), making saturation concerns minimal for typical GC observations at distances $\gtrsim 1$ kpc.

\subsubsection{Angular Resolution and Source Resolvability}

The high angular resolution of ELT/MICADO is crucial for detecting SL sources in crowded GC cores. Its high-resolution mode has a pixel scale of 1.5 mas (Davies \etal 2021), while our calculations adopt a practical 12 mas source-separation threshold in the $K$ filter (R. Davies, private comm.) to allow for atmospheric turbulence, adaptive-optics performance variations, and photometric-precision requirements.

This angular resolution translates to linear resolutions that vary dramatically with cluster distance. At our reference distances of 1 kpc, 10 kpc, and 100 kpc, the 12 mas resolution limit corresponds to projected linear separations of $5.8 \times 10^{-5}$ pc, $5.8 \times 10^{-4}$ pc, and $5.8 \times 10^{-3}$ pc, respectively.

To assess the impact of crowding on source resolvability, we employed the COCOA code (Askar \etal 2018) to project our three-dimensional cluster models onto the observational plane. Minimum projected stellar separations were then compared with the physical separation corresponding to the adopted angular threshold at each GC distance. This analysis accounts for the spatial distributions and density profiles of evolved GCs, including mass segregation and core collapse (Table~\tabresults). Details of the projection algorithm are given by Askar \etal (2018).

\begin{table}
\centering
\caption{Fraction of self-lensing sources resolvable by ELT/MICADO at different cluster distances (1, 10, or $100\kpc$), accounting for crowding effects in dense stellar environments. Values represent the mean fraction across cluster snapshots with 1-$\sigma$ uncertainties (given in parentheses as errors in the last significant digit(s)).}
\footnotesize
\begin{tabular}{llll}
\toprule
 & \multicolumn{3}{c}{Fraction of Resolvable Sources} \\
Label & 1 kpc & 10 kpc & 100 kpc \\
\midrule
npop1-fb10-nTF & 1.00 & 0.99(1) & 0.46(3) \\
npop1-fb95-nTF & 1.00 & 0.98(1) & 0.37(2) \\
npop2-cpop05-fb10-TF & 1.00 & 1.00 & 0.86(5) \\
npop2-cpop05-fb10-TF-NF & 1.00 & 1.00(1) & 0.82(5) \\
npop2-cpop05-fb10-nTF & 1.00 & 0.99(1) & 0.29(3) \\
npop2-cpop05-fb10-nTF-NF & 1.00 & 0.98(2) & 0.37(8) \\
npop2-cpop05-fb95-TF & 1.00 & 0.99 & 0.67(4) \\
npop2-cpop05-fb95-TF-NF & 1.00 & 1.00 & 0.82(2) \\
npop2-cpop05-fb95-nTF & 1.00 & 0.98(1) & 0.34(1) \\
npop2-cpop05-fb95-nTF-NF & 1.00 & 0.99(1) & 0.38(4) \\
npop2-cpop2-fb10-TF-NF & 1.00 & 1.00 & 0.88(3) \\
npop2-cpop2-fb10-nTF & 1.00 & 0.97(1) & 0.15(4) \\
npop2-cpop2-fb10-nTF-NF & 0.99(1) & 0.91(7) & 0.23(10) \\
npop2-cpop2-fb95-TF-NF & 1.00 & 1.00 & 0.88(1) \\
npop2-cpop2-fb95-nTF & 1.00 & 0.98(1) & 0.19(1) \\
npop2-cpop2-fb95-nTF-NF & 1.00 & 0.90(2) & 0.11(2) \\
\bottomrule
\end{tabular}

\end{table}

Table~\tabeltaccessibility presents the fraction of SL sources that remain resolvable at each distance after accounting for crowding effects. At nearby distances (1 kpc), essentially all SL sources remain resolvable across all simulation configurations. At intermediate distances (10 kpc), representing the median distance of the Galactic GCs, resolvability remains high ($\gtrsim 85\%$) for most configurations. The modest reductions observed in some simulations reflect the onset of crowding effects in the densest cluster regions, particularly for nTF which form centrally concentrated clusters. At largest distances (100 kpc), TF maintain higher resolvability ($\sim 66-89\%$) compared to nTF systems ($\sim 10-48\%$), reflecting the larger physical separations and reduced crowding susceptibility of wide binary systems.

\subsubsection{Observational Probability and Detection Statistics}\label{sec:obsprob}

A critical consideration for SL detection programs is the probability that a source will be observed during its magnification phase. Unlike microlensing events, which are typically single, unrepeated phenomena, SL sources exhibit periodic variability tied to the orbital motion of the binary system.

The probability that a single exposure overlaps an SL magnification phase is approximated by:
\begin{equation} 
\pobs = \min\left(1,\frac{\taueff + \texp}{\porb}\right)
\end{equation}
where $\taueff$ is the effective Einstein ring crossing time, $\texp$ is the exposure duration, and $\porb$ is the orbital period of the binary system. \new{For the adopted 1-minute benchmark visits, we assessed exposure-time dilution using $f_{\rm dil}=\min(1,\taueff/\texp)$. Its $\Erawa$-weighted mean is 0.999852; only 0.061\% of sources have $f_{\rm dil}<1$. We therefore do not apply a dilution correction to the reported benchmark rates.}

\begin{figure}
\centering
\includegraphics[width=\linewidth]{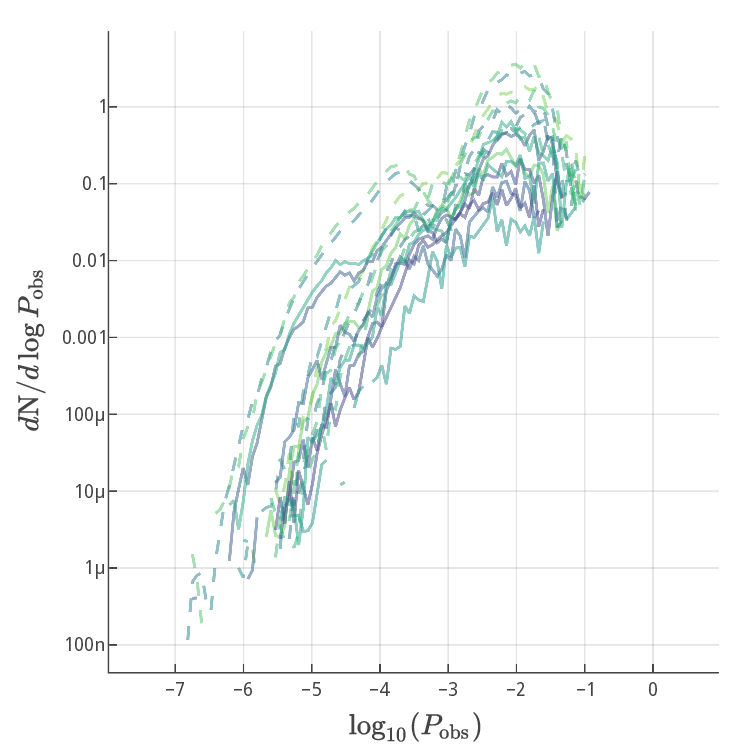}
\FigCap{Distribution of single-exposure phase-overlap probabilities for self-lensing sources across all simulation configurations, calculated for $\texp=1$ min. Labels as in Figure~\figmdistribution.}
\end{figure}

Figure~\figPobsdistribution shows the distribution of single-exposure phase-overlap probabilities across all simulated SL systems. The distribution exhibits a broad peak around $\log_{10} \pobs \approx -1.5$, corresponding to an overlap probability of $\sim 3\%$ per exposure. The distribution extends to very low probabilities ($\log_{10} \pobs \approx -6$), reflecting systems with very long orbital periods.

\subsubsection{Recurrent observations}

The periodic nature of SL enables significant improvements in detection efficiency through multiple observations. Recurrent observations are particularly important for distinguishing SL from microlensing events \new{(\eg Paczynski 1986)} and other transient phenomena like stellar flares \new{(\eg Kowalski 2024)}, as the periodic signature provides definitive identification.

For SL surveys, we define the expected coverage as the number of data points obtained during a single lensing event:
\begin{equation}
\Ecov= \max\left(1,\frac{\taueff + \texp}{\tcadence}\right)
\end{equation}
where $\tcadence$ is the time interval between successive observations. The expected number of recurrences during a survey of duration $\tsurvey$ is:
\begin{equation}
\Erec = \frac{\tsurvey}{\porb}\min\left(1,\frac{\taueff + \texp}{\tcadence}\right)
\end{equation}

These prescriptions agree with the "brute force" method described in Wiktorowicz \etal (2021) for $\tcadence > \taueff$ and include the exposure time correction.

Table~\tabsurveystrategy presents the fraction of SL sources meeting various coverage and recurrence criteria for different survey strategies. Daily cadence provides the best performance. \new{For 1-year surveys, the median fractions with more than one and more than ten recurrences are 86\% and 11\%, respectively; for 10-year surveys they increase to 98\% and 86\%.} However, few sources ($\ll 1\%$) achieve high coverage ($>10$ data points per event).

\new{At a 10-day cadence, the median fractions with more than one recurrence are 11\% and 87\% for 1-year and 10-year surveys, while the corresponding fractions with more than ten recurrences are $\sim0\%$ and 11\%.}

The total expected number of observational data points for each source during an SL event is given by:
\begin{equation}
\Etot = \Ecov \times \Erec
\end{equation}

\begin{figure*}
    \centering
    \includegraphics[width=\linewidth]{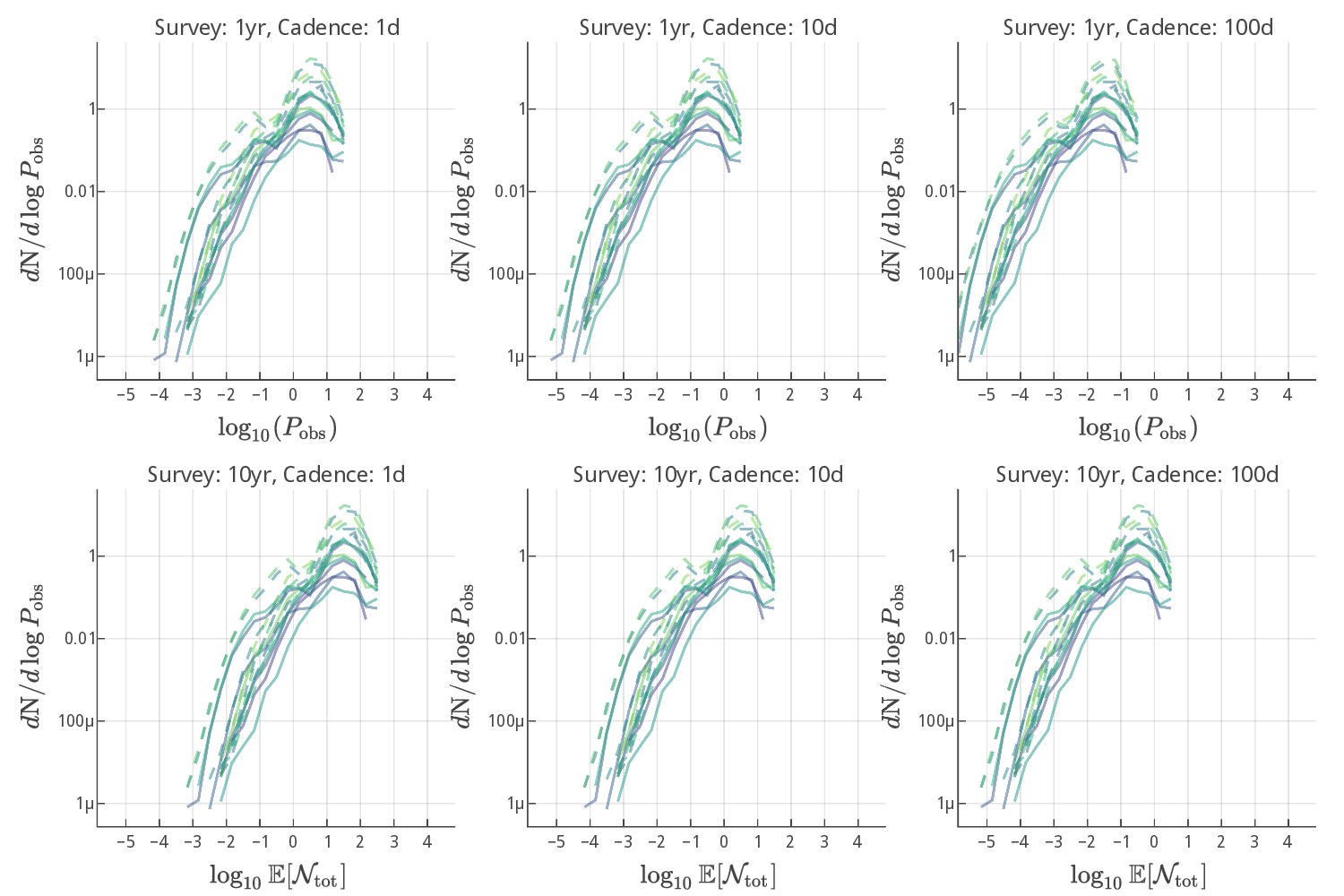}
    \FigCap{Distributions of total expected data points per self-lensing source for different survey strategies. Top panels show results for 1-year surveys, bottom panels for 10-year surveys. Columns represent observing cadences of 1, 10, and 100 days. The exposure duration is 1 minute. Labels as in Figure~\figmdistribution.}
\end{figure*}

Figure~\figEtotdistributions presents the distribution of total expected data points across different survey strategies. The results highlight several key findings:
\begin{itemize}
    \item Extending surveys from 1 to 10 years dramatically increases the number of data points, particularly for the tail of well-observed sources. This improvement is prominent for all cadence values.
    \item Daily observations provide an order of magnitude more data points than a 10-day cadence, which is an order of magnitude more effective than a 100-day cadence.
\end{itemize}

\subsubsection{Photometric Precision}

The detection of SL events ultimately depends on achieving sufficient photometric precision to measure the small magnitude variations. For AO observations of crowded stellar fields, relative photometric precision is strongly affected by uncertainties in point-spread function (PSF) characterization. Lu \etal (2010) showed that Keck can achieve relative photometric accuracies of $\sim3\%$ ($\sim33$ mmag) in crowded fields. ELT is a much larger (39 m) telescope with the MORFEO AO system (Ciliegi \etal 2024), which features advanced PSF reconstruction capabilities. \new{We adopt 10 mmag as an optimistic per-epoch photometric scatter (R. Davies, priv. comm.) and 33 mmag as a conservative value (Lu \etal 2010). We apply these magnitude-independent per-epoch scatter values to all sources passing the single-visit magnitude cut and do not model the increase in photon or background noise toward this limit. The detectable fractions and rates in Tables 3 and 5 should therefore be regarded as optimistic upper limits; including magnitude-dependent noise would reduce the predicted yields further.}

For multi-epoch observations, the effective noise level decreases with the square root of the number of measurements:
\begin{equation}
\sigma_{\rm noise,eff} = \frac{\sigmanoise}{\sqrt{\Etot}}.
\end{equation}
This scaling assumes statistically independent, uncorrelated per-epoch errors. Time-correlated AO or PSF systematics would not average down in this way, so the resulting multi-epoch rates are optimistic in this respect.

We define a SL event as detectable when the magnification exceeds 3 times the effective photometric uncertainty:
\begin{equation}
|\Delta m| > 3 \sigma_{\rm noise,eff}.
\end{equation}

We emphasize that this 3-$\sigma$ criterion is an illustrative threshold for targeted observations of systems with known or constrained orbital periods. Blind searches would require higher thresholds (\eg S/N$>\sim6.3$ in Gilliland \etal 2000 or S/N$>10$ in Yamaguchi \etal 2024). Periodic phase folding can reduce false-alarm probabilities relative to an isolated measurement, but a full search analysis is beyond our scope.

\begin{table}
\caption{Fraction of self-lensing sources producing magnification events detectable above the photometric noise threshold for a 10-year survey with 1-day cadence. This diagnostic assumes independent per-epoch errors and is not the 1-year configuration used in Table~\tabcomprehensivepredictions. Values represent means across snapshots with 1-$\sigma$ uncertainties (given in parentheses as errors in the last significant digit(s)). We show results for $\sigmanoise=10$ mmag (our optimistic assumption) and a conservative value of $\sigmanoise=33$ mmag (Lu \etal 2010).}
    \centering
\begin{tabular}{lll}
\toprule
     & \multicolumn{2}{c}{Fraction of Systems} \\
     &  \multicolumn{2}{c}{Above The Noise} \\
     & \multicolumn{2}{c}{ for $\sigmanoise=$} \\
Label & $10$ mmag & $33$ mmag \\
\midrule
npop1-fb10-nTF & 0.058(4) & 0.025(2) \\
npop1-fb95-nTF & 0.026(3) & 0.011 \\
npop2-cpop05-fb10-TF & 0.068(11) & 0.042(6) \\
npop2-cpop05-fb10-TF-NF & 0.045(3) & 0.027(2) \\
npop2-cpop05-fb10-nTF & 0.050(2) & 0.018(1) \\
npop2-cpop05-fb10-nTF-NF & 0.074(10) & 0.020(3) \\
npop2-cpop05-fb95-TF & 0.036(2) & 0.016(1) \\
npop2-cpop05-fb95-TF-NF & 0.033(2) & 0.017(1) \\
npop2-cpop05-fb95-nTF & 0.024(1) & 0.0080(3) \\
npop2-cpop05-fb95-nTF-NF & 0.021(1) & 0.0055(5) \\
npop2-cpop2-fb10-TF-NF & 0.046(2) & 0.028(1) \\
npop2-cpop2-fb10-nTF & 0.061(7) & 0.020(1) \\
npop2-cpop2-fb10-nTF-NF & 0.085(17) & 0.036(9) \\
npop2-cpop2-fb95-TF-NF & 0.031(1) & 0.015(1) \\
npop2-cpop2-fb95-nTF & 0.024(2) & 0.0079(6) \\
npop2-cpop2-fb95-nTF-NF & 0.054(6) & 0.040(3) \\
\bottomrule
\end{tabular}

\end{table}

\new{Table~\tabphotometricdetection presents the detectable fractions for the optimistic 10 mmag and conservative 33 mmag per-epoch scatter. The fractions span $2.1$--$8.5\%$ for 10 mmag and $0.55$--$4.2\%$ for 33 mmag. Low-initial-binary-fraction models often yield higher detectable fractions, although this trend is not universal.}

\subsubsection{Comprehensive Predictions}\label{sec:conservativepredictions}

\new{To provide illustrative estimates under the adopted assumptions, we combine the single-visit magnitude limit, angular resolution, photometric scatter, and temporal sampling described above.} Table~\tabcomprehensivepredictions presents a monitoring benchmark and a random-epoch geometric baseline.

The monitoring benchmark represents 1 year of daily \new{$\texp = 1$ min visits in \Ks-band and assumes independent per-epoch errors. The comparison quantity is the random-epoch geometric probability that a given SL source is in its magnification phase during one randomly timed visit. It is not an observational detection rate: identifying an SL binary requires multiple epochs to establish periodicity.}

\new{The monitoring benchmark yields rates roughly two to three orders of magnitude above the random-epoch geometric baseline because it combines repeated temporal coverage with the assumed statistical improvement in photometric precision. For the 10 mmag scatter, whole-cluster rates span $2.62\times10^{-3}$--$7.57\times10^{-2}$ at 1 kpc, $2.2\times10^{-6}$--$1.409\times10^{-2}$ at 10 kpc, and zero--$1.4\times10^{-7}$ at 100 kpc. For the conservative 33 mmag scatter, the corresponding ranges are $8.7\times10^{-5}$--$2.65\times10^{-2}$, $1.6\times10^{-7}$--$4.10\times10^{-4}$, and zero at 100 kpc. The strong decline with distance reflects the 1-minute magnitude limit and increasing crowding. No single trend with initial binary fraction or tidal filling describes all models.}

Figure~\figELTdistributions shows the distributions of $\taueff$, \Ks magnitude, and $\porb$ for the ELT-detectable subset in the 10 kpc, 10 mmag monitoring benchmark. The $\taueff$ values are shifted toward longer durations than in the full population, and the detectable systems preferentially occupy brighter magnitudes and wider orbits.

Figure~\figAllELTdistributions compares key properties of the full SL population (thin lines) with the ELT-detectable subset (thick coloured lines). The detectable systems exhibit selection effects. Orbital periods shift toward longer values, with the detectable population peaking at $\porb\sim 3$ yr compared to $\sim$ days for the general population, and extending to very wide systems ($\porb\sim200$ kyr). This bias reflects the magnitude-limited selection, which favours systems with larger Einstein radii and higher magnifications.

Lens-mass distributions have similar shapes for the general and ELT-detectable populations. Among WD types, carbon-oxygen WDs dominate both samples. NSs and BHs remain rare in the detectable population because of their low abundance in 12 Gyr clusters and dynamical ejections.

\section{Discussion}

\subsection{Physical Constraints and Selection Effects}

The detection of SL events in GCs faces several fundamental challenges that limit observable populations and constrain system parameters. Dynamical disruption represents perhaps the most significant limitation for wide, highly magnified SL sources. The dense stellar environments characteristic of GC cores (stellar densities $\sim 10^3-10^6$ stars pc$^{-3}$; Baumgardt \& Hilker 2018) subject wide binaries to frequent gravitational perturbations that can either harden close systems or completely dissolve wider ones (Heggie 1975).

The stellar evolutionary constraints in old GC environments further restrict detectable SL populations. Unlike young stellar systems where massive MS stars provide bright sources, 12 Gyr-old GCs are dominated by low-mass MS stars ($M \lesssim 0.8\msun$, i.e. the turn-off mass for the considered clusters) and evolved stellar remnants. Massive stars that could serve as luminous sources have already evolved off the MS and either formed compact objects or been ejected from the cluster through dynamical interactions or supernova kicks (Hurley \etal 2007).

The finite-size effects of WD lenses introduce additional complexity through the intrinsic degeneracy between lensing-induced magnification and occultation-induced demagnification (Han 2016). When the physical radius of a WD lens becomes comparable to its Einstein radius ($\rL = \RL/\RE \sim 1$), occultation effects can significantly reduce the total magnification and, in extreme cases, cause demagnification below baseline levels. This effect is particularly pronounced for close binary systems with low-mass WD lenses, where "deep eclipses" dominate over the lensing signal (Sajadian 2025).

\subsection{Comparison with Galactic Field Populations}

This study represents the first dedicated exploration of SL in GC environments, complementing previous investigations of SL in Galactic field populations (Wiktorowicz \etal 2021, Wiktorowicz \etal 2025b). The environmental differences between GCs and the Galactic field lead to distinct SL populations and observational characteristics.

The relationship between GC-based and field-based SL predictions is complex due to the interconnected evolutionary history of these stellar populations. Previous population synthesis studies of Galactic field environments were primarily based on isolated binary evolution models and were observationally limited to the nearby galactic disk, not encompassing the full GC population. However, GCs continuously eject stellar systems throughout their evolution that subsequently mix into the field populations. This evolutionary mixing means that current field population models likely already incorporate some contribution from former GC systems, though disentangling these populations among currently observed systems remains challenging with available observational tools.

Galactic field environments are characterized by diverse metallicities, ages, and star formation histories, leading to SL populations dominated by wide, non-interacting binaries with NS and BH lenses. The field studies predict $\sim 100-10,000$ detectable SL events with NS/BH lenses for wide-field surveys like ZTF and LSST (Wiktorowicz \etal 2021), far exceeding our GC-specific yields due to the vastly larger surveyed volume, smaller distances and younger stellar populations.

In contrast, GC environments offer several unique advantages for SL studies. All stars within a given cluster share essentially identical distances and ages, eliminating degeneracies that complicate field population studies. The high stellar densities and old ages of GCs ($\sim 12$ Gyr) result in enhanced populations of evolved binaries, including substantial numbers of double WD systems that remain individually undetectable through conventional methods (Hellstr{\"o}m \etal 2024). While WD lenses produce lower magnification amplitudes due to occultation effects, they offer potential advantages including occultation signatures for system characterization and represent the most abundant compact object population in old stellar systems.

Double WD systems represent a particularly interesting subset of our predicted SL population. These systems can exhibit mutual lensing events where either component can act as lens or source, depending on orbital phase. Such configurations offer unique opportunities to determine individual component masses and radii through detailed light-curve modelling (Agol 2002), providing direct tests of WD cooling models and mass-radius relations in the low-luminosity regime.

The spatial distribution of SL sources, with 55--93\% residing within the half-mass radius (Table~\tabresults), reflects the effects of mass segregation in GCs. Massive stellar remnants that form lenses migrate toward cluster centres through dynamical friction, concentrating detectable systems in the highest-density regions. This spatial bias has implications for both observational strategy (prioritizing core observations) and detectability (crowding in GC cores).

\subsection{Limitations and Future Improvements}

Several limitations in our current analysis warrant future refinement. The uniform source brightness approximation neglects limb-darkening effects that can influence magnification profiles, particularly for evolved stellar sources with significant radial brightness variations (Claret 2000, Sajadian \& Afshordi 2024). More sophisticated modelling incorporating stellar atmosphere models would improve system parameter determination but is computationally intensive for survey-level predictions.

Our detectability criteria focus on geometric and photometric considerations without detailed treatment of systematic errors, atmospheric effects, or instrumental complications. Real observations will face additional challenges from telescope guiding errors, PSF variations, and contamination from nearby sources that could affect photometric precision. The optimistic 10 mmag and conservative 33 mmag per-epoch photometric scatters bracket some of this uncertainty, but actual performance may be poorer in the most crowded fields.

The MOCCA simulations, while sophisticated, employ several approximations in stellar evolution and dynamical modelling that could influence SL predictions. Additionally, the current study would benefit from more diverse models exploring different metallicities and initial cluster properties, as variations in these parameters can significantly affect the formation and evolution of compact binaries. Ongoing updates to stellar evolution and dynamical modules, combined with this expanded parameter space exploration, may alter the predicted populations of evolved binaries and their SL characteristics.

\new{Additionally, predictions tailored to specific well-studied GCs (e.g. M4, 47 Tuc, NGC 6397) would require dedicated MOCCA simulations with carefully calibrated initial conditions matching the observed mass, concentration, binary fraction, and dynamical history of each target. Such cluster-specific modelling represents a natural and important extension of this work, but constitutes material for separate follow-up studies.}

Our detectability criteria focus on photometric signal-to-noise without explicit treatment of false alarm rates from astrophysical sources (stellar variability, flares) or statistical trials factors. Actual observing programs would need to adopt higher significance thresholds for blind searches, though the strictly periodic nature of SL provides strong discrimination against random false positives. The detection rates presented here should therefore be considered optimistic upper limits for blind searches, though more realistic for targeted monitoring of systems with known or suspected orbital periods derived from radial velocity or astrometric observations.

\section{Conclusions}

This work presents a theoretical investigation of SL phenomena in GC environments and a framework for assessing their potential observability with ELT/MICADO.

Our MOCCA simulations predict that present-day GCs contain $\sim1$--$50$ SL sources with magnifications ($\musl > 1$), with yields strongly dependent on initial GC properties. Systems with high initial binary fractions ($\fracb = 95\%$) consistently outperform low binary fraction configurations ($\fracb = 10\%$). Tidally filling clusters with new features show the highest predicted magnifications, often exceeding $\musl > 100$, though highly magnified sources remain rare ($\lesssim 0.1$ systems per cluster with $\musl > 2$).

The predicted SL populations exhibit several distinctive characteristics. WD lenses dominate the sample, paired predominantly with low-mass MS companions, reflecting the evolved nature of $12 \gyr$-old stellar systems. The apparent magnitude distributions show characteristic bimodal structure with peaks at $m \approx 24$ and $32$ mag (at 10 kpc), indicating populations accessible to both current facilities and next-generation telescopes. Temporal signatures are characterized by short crossing times ($\taueff \sim 2$ hours for typical systems), demanding high-cadence observations for optimal identification.

\new{We use MICADO's anticipated angular resolution and sensitivity to estimate which simulated SL sources satisfy the adopted detectability criteria. For the idealized 1-year daily-cadence benchmark, the maximum whole-cluster expectations are approximately 0.076 and 0.027 sources per cluster for the optimistic 10 mmag and conservative 33 mmag scatters, respectively. These maxima occur at the illustrative 1 kpc distance and are upper limits for a single MICADO pointing, not predictions for a known cluster or a fully specified observing programme.

The periodic nature of SL events provides significant advantages for systematic monitoring. Multi-year campaigns with daily cadence provide greater temporal coverage and statistical precision than the random-epoch geometric baseline.

Our results illustrate how the modeled rates depend on cluster distance, cadence, and photometric scatter. Nearby clusters generally give higher whole-cluster rates, while the model-to-model spread motivates future target-specific calculations. We treat 10 mmag photometric scatter as optimistic and 33 mmag as a conservative comparison.}

DWD systems, representing $\sim1\%$ of predicted sources, offer opportunities for mutual lensing studies that can constrain individual component properties through detailed light-curve modelling (Sajadian 2025). More broadly, the predicted SL populations encode information about cluster dynamics through their spatial distribution, with $\sim 55$--$93\%$ of systems concentrated within cluster half-mass radii, providing direct probes of mass segregation and dynamical evolution in dense stellar environments.

Future improvements to this analysis should incorporate refined stellar atmosphere models to account for limb-darkening effects, enhanced treatment of systematic errors in crowded-field photometry, and updated stellar evolution prescriptions as they become available. Coordinated observations across multiple facilities -- leveraging wide-field surveys for statistical populations, high-resolution follow-up for detailed characterization, and space-based platforms for optimal temporal coverage -- will maximize the scientific return from SL studies.

The methodology developed here provides a framework for SL predictions in other dense stellar environments, including nuclear star clusters, young massive clusters, and other galaxies. Future SL observations could constrain otherwise difficult-to-detect compact-object binaries in dense stellar systems.

\Acknow{
We are thankful to Arkadiusz Hypki for the help with code development. GW was supported by the Polish National Science Center (NCN) through the grant 2021/41/B/ST9/01191. MM acknowledges support from STFC small awards (ST/Y001699/1). AI acknowledges support from the Royal Society. AA acknowledges that this research was funded in part by the Polish National Science Center (NCN) grant number 2024/55/D/ST9/02585. For the purpose of Open Access, the author has applied a CC BY public copyright licence to any Author Accepted Manuscript (AAM) version arising from this submission.
}





\appendix

\begin{figure*}
\section{Additional figures}
    \centering
    \includegraphics[width=\linewidth]{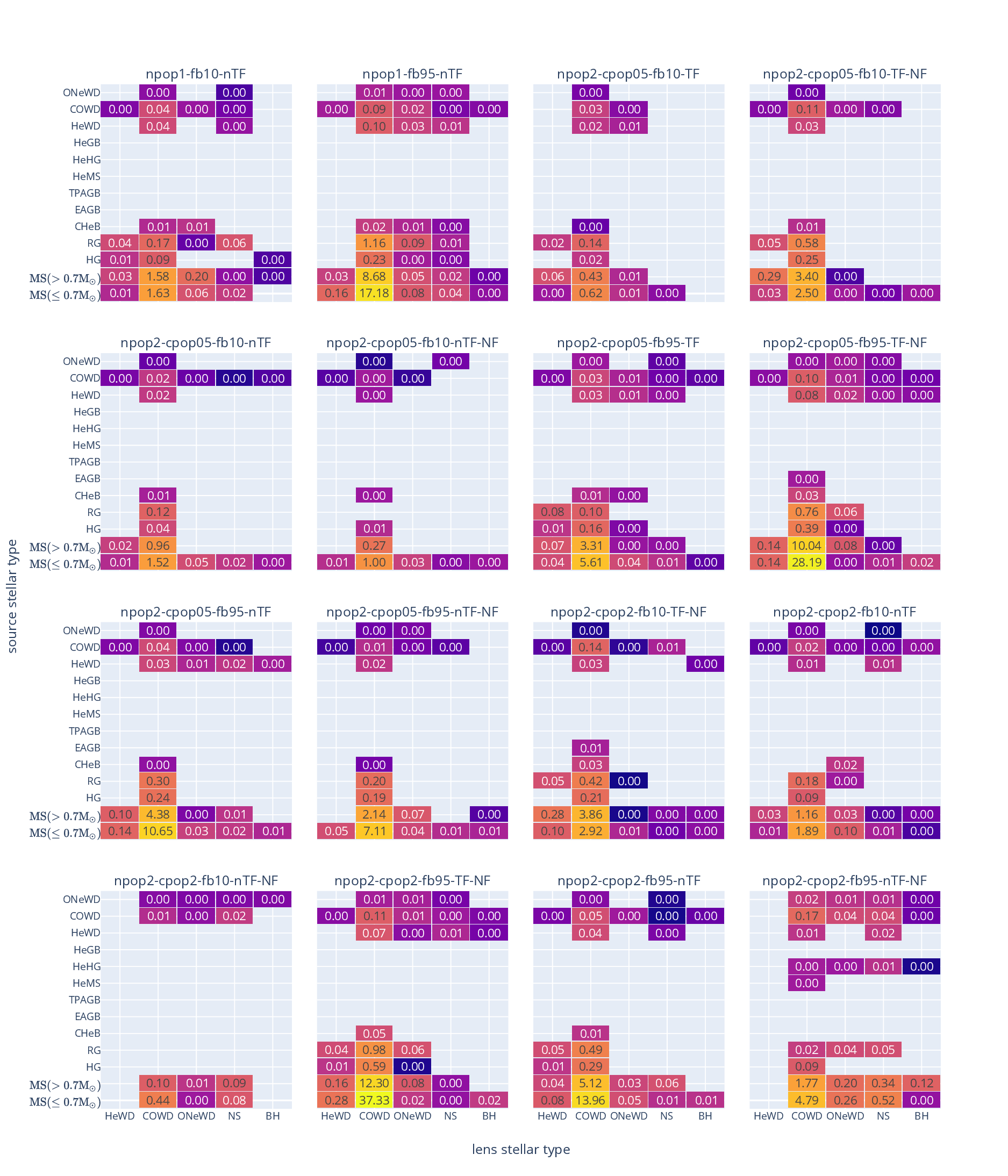}
    \FigCap{Distribution of lens and source stellar evolutionary types. Only simulations with non-zero predictions are shown. Numbers represent expected numbers ($\Erawa$). For $\Erawa<0.01$ only $0.00$ is shown. Stellar types:MS - main sequence, HG - Hertzsprung gap, RG - red giant, CHeB - core helium burning, EAGB - early asymptotic giant branch, TPAGB - thermally pulsating asymptotic giant branch, HeGB - helium giant branch, HeWD - helium white dwarf, COWD - carbon-oxygen white dwarf, ONeWD - oxygen-neon white dwarf, NS - neutron star, BH - black hole.}
\end{figure*}

\begin{figure*}
    \centering
    \includegraphics[width=\linewidth]{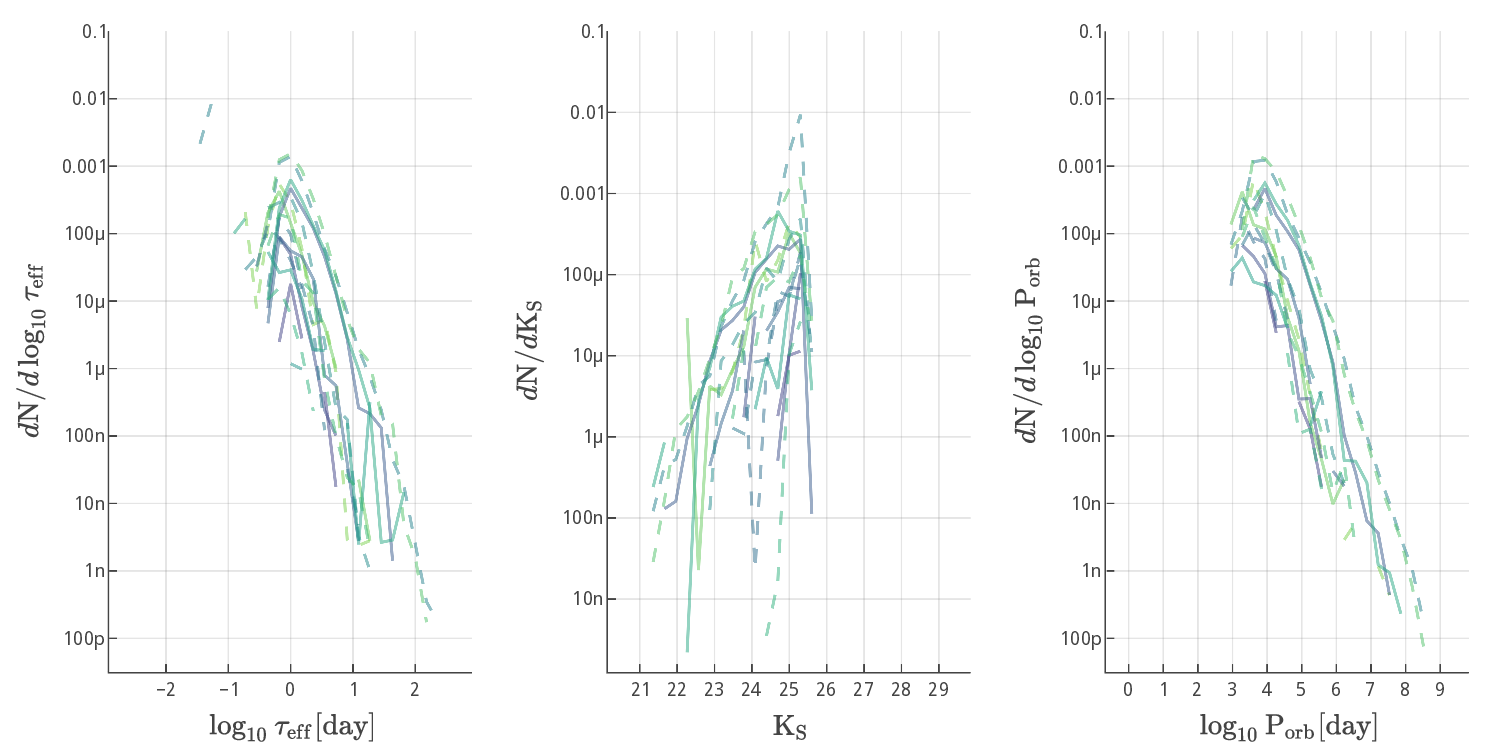}
    \FigCap{Distributions of the effective Einstein ring crossing time ($\taueff$; left), magnitude in \Ks filter (middle), and orbital period ($\porb$; right) for the ELT-observable subset at 10 kpc in the 10 mmag monitoring benchmark. Labels as in Figure~\figmdistribution.}
\end{figure*}

\begin{figure*}
    \centering
    \includegraphics[width=\linewidth]{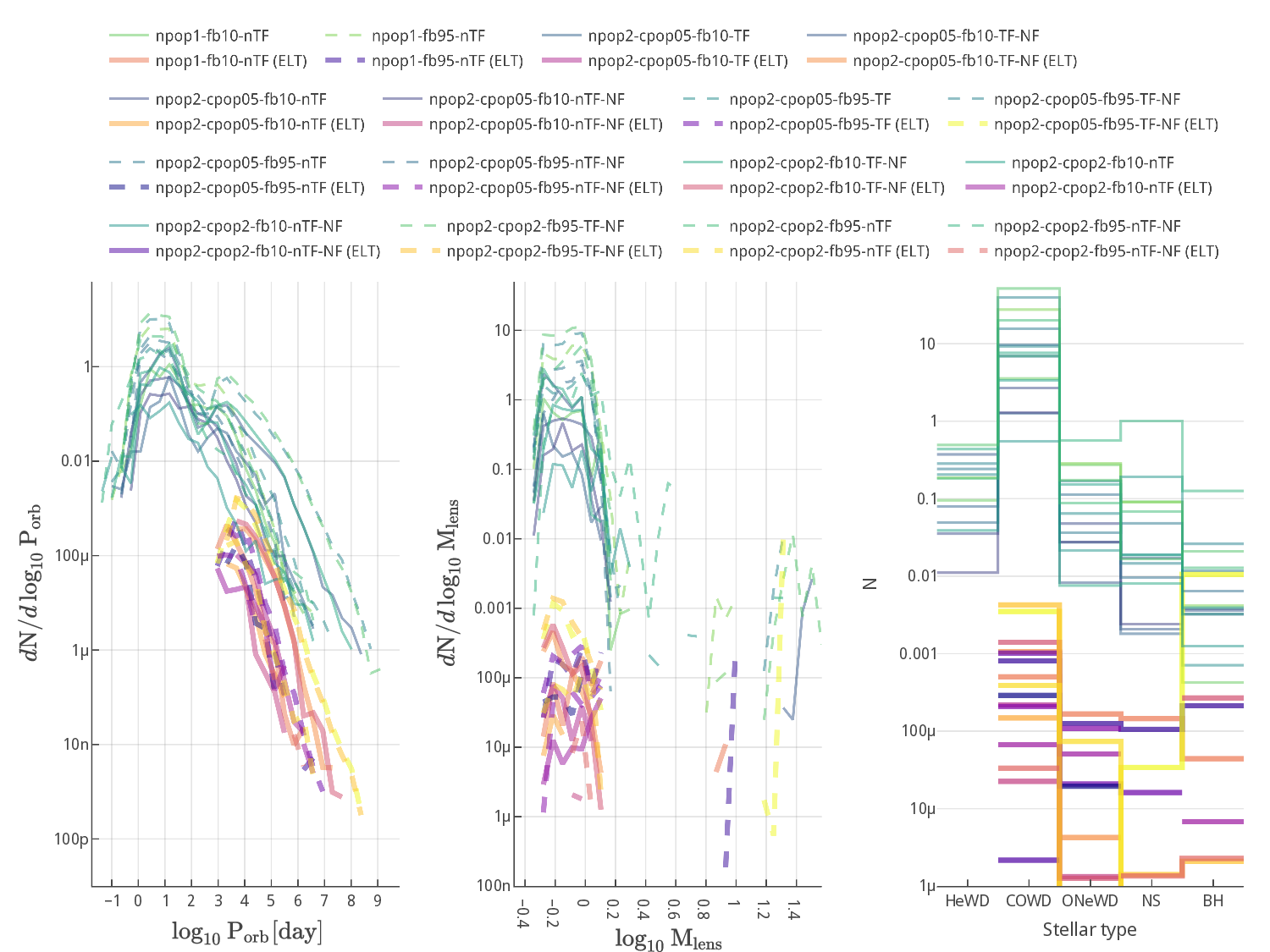}
    \FigCap{Distributions of orbital periods (left), lens masses (middle), and lens types (right) for the full SL population (thin green-blue lines) and the ELT-detectable subset at 10 kpc in the 10 mmag monitoring benchmark (thick, coloured). Presented stellar types represent: HeWD - Helium White Dwarf, COWD - Carbon-Oxygen White Dwarf; ONeWD - Oxygen-Neon White Dwarf; NS - Neutron Star; BH - Black Hole.}
\end{figure*}

\begin{sidewaystable*}
\section{Additional tables}
\caption{Fraction of self-lensing sources satisfying coverage and recurrence criteria for different survey strategies. Coverage fractions ($\Ecov$) represent sources with multiple data points per lensing event, while recurrence fractions ($\Erec$) indicate sources whose expected recurrence count exceeds the stated threshold. Results are shown as minimum, median, and maximum values across all simulations for 1-year and 10-year survey durations with daily, 10-day, and 100-day cadences.}
    \centering
    \begin{tabular}{llrrrrrrrrrrrr}
    \toprule
     &  & \multicolumn{3}{c}{$\Ecov>1$} & \multicolumn{3}{c}{$\Ecov>10$} & \multicolumn{3}{c}{$\Erec>1$} & \multicolumn{3}{c}{$\Erec>10$} \\
    $\tsurvey$ [yr]& $\tcadence$ [days] & min & median & max & min & median & max & min & median & max & min & median & max \\
    \midrule
    \multirow[t]{3}{*}{1} & 1 & 9.7e-05 & 0.012 & 0.066 & 0 & 0.00011 & 0.0053 & 0.67 & 0.86 & 0.90 & 0.021 & 0.11 & 0.21 \\
     & 10 & 0 & 0.00011 & 0.0053 & 0 & 0 & 0.0011 & 0.021 & 0.11 & 0.21 & 0 & 0 & 0 \\
     & 100 & 0 & 0 & 0.0011 & 0 & 0 & 6.1e-09 & 0 & 0 & 0 & 0 & 0 & 0 \\
    \cline{1-14}
    \multirow[t]{3}{*}{10} & 1 & 9.7e-05 & 0.012 & 0.066 & 0 & 0.00011 & 0.0053 & 0.95 & 0.98 & 0.99 & 0.67 & 0.86 & 0.90 \\
     & 10 & 0 & 0.00011 & 0.0053 & 0 & 0 & 0.0011 & 0.68 & 0.87 & 0.90 & 0.021 & 0.11 & 0.21 \\
     & 100 & 0 & 0 & 0.0011 & 0 & 0 & 6.1e-09 & 0.021 & 0.11 & 0.21 & 0 & 0 & 0 \\
    \cline{1-14}
    \bottomrule
    \end{tabular}

\end{sidewaystable*}

\newcommand{\rates}[2]{\shortstack{$#1$\\$#2$}}
\begin{sidewaystable*}
    \caption{Comprehensive ELT self-lensing benchmarks under the adopted idealized cuts. The monitoring columns assume 1 year of daily 1-minute visits in \Ks-band; the random-epoch columns give the expected number passing the observational cuts and magnified during one randomly timed visit, not identification from a single image. Each entry gives the result for 10 mmag (top) and 33 mmag (bottom) independent per-epoch scatter. Values use the \new{\Ks $=25.46$ mag single-visit limit} and a 12 mas source-separation threshold; no photon-noise or dilution correction is applied. Parentheses give the snapshot-to-snapshot standard deviation in the last significant digit(s). Rates refer to the whole cluster and are upper limits for a single MICADO pointing (Section~\ref{sec:ELT}).}
\centering
\scriptsize
\begin{tabular}{lllllll}
\toprule
 & \multicolumn{2}{c}{D=1 kpc} & \multicolumn{2}{c}{D=10 kpc} & \multicolumn{2}{c}{D=100 kpc} \\
Label & monitoring & random epoch & monitoring & random epoch & monitoring & random epoch \\
\midrule
npop1-fb10-nTF & \rates{9.13(48)\times10^{-3}}{1.268(83)\times10^{-3}} & \rates{2.50(13)\times10^{-5}}{3.48(23)\times10^{-6}} & \rates{8.6(10)\times10^{-4}}{5.0(53)\times10^{-5}} & \rates{2.35(29)\times10^{-6}}{1.4(15)\times10^{-7}} & \rates{0}{0} & \rates{0}{0} \\
npop1-fb95-nTF & \rates{3.23(24)\times10^{-2}}{4.20(28)\times10^{-3}} & \rates{8.86(65)\times10^{-5}}{1.151(78)\times10^{-5}} & \rates{1.25(44)\times10^{-3}}{1.42(80)\times10^{-5}} & \rates{3.4(12)\times10^{-6}}{3.9(22)\times10^{-8}} & \rates{0}{0} & \rates{0}{0} \\
npop2-cpop05-fb10-TF & \rates{2.62(16)\times10^{-3}}{5.31(20)\times10^{-4}} & \rates{7.19(44)\times10^{-6}}{1.455(54)\times10^{-6}} & \rates{2.18(21)\times10^{-4}}{1.43(41)\times10^{-5}} & \rates{5.96(58)\times10^{-7}}{3.9(11)\times10^{-8}} & \rates{0}{0} & \rates{0}{0} \\
npop2-cpop05-fb10-TF-NF & \rates{1.10(30)\times10^{-2}}{4.4(32)\times10^{-3}} & \rates{3.2(11)\times10^{-5}}{1.4(11)\times10^{-5}} & \rates{1.070(38)\times10^{-3}}{1.09(12)\times10^{-4}} & \rates{2.93(10)\times10^{-6}}{2.98(32)\times10^{-7}} & \rates{0}{0} & \rates{0}{0} \\
npop2-cpop05-fb10-nTF & \rates{8.7(10)\times10^{-3}}{1.36(69)\times10^{-3}} & \rates{2.38(28)\times10^{-5}}{3.7(19)\times10^{-6}} & \rates{1.48(21)\times10^{-4}}{3.0(19)\times10^{-6}} & \rates{4.05(57)\times10^{-7}}{8.3(52)\times10^{-9}} & \rates{0}{0} & \rates{0}{0} \\
npop2-cpop05-fb10-nTF-NF & \rates{9.29(63)\times10^{-3}}{1.61(52)\times10^{-3}} & \rates{2.55(17)\times10^{-5}}{4.4(14)\times10^{-6}} & \rates{2.38(88)\times10^{-5}}{2.3(26)\times10^{-6}} & \rates{6.5(24)\times10^{-8}}{6.3(73)\times10^{-9}} & \rates{0}{0} & \rates{0}{0} \\
npop2-cpop05-fb95-TF & \rates{1.389(47)\times10^{-2}}{2.589(57)\times10^{-3}} & \rates{3.81(13)\times10^{-5}}{7.09(16)\times10^{-6}} & \rates{1.054(41)\times10^{-3}}{4.2(13)\times10^{-5}} & \rates{2.89(11)\times10^{-6}}{1.15(36)\times10^{-7}} & \rates{0}{0} & \rates{0}{0} \\
npop2-cpop05-fb95-TF-NF & \rates{7.23(42)\times10^{-2}}{2.28(20)\times10^{-2}} & \rates{2.38(28)\times10^{-4}}{9.1(21)\times10^{-5}} & \rates{1.409(32)\times10^{-2}}{3.232(75)\times10^{-4}} & \rates{4.97(28)\times10^{-5}}{8.86(21)\times10^{-7}} & \rates{0}{0} & \rates{0}{0} \\
npop2-cpop05-fb95-nTF & \rates{2.23(47)\times10^{-2}}{4.4(50)\times10^{-3}} & \rates{6.2(16)\times10^{-5}}{1.3(16)\times10^{-5}} & \rates{3.09(66)\times10^{-4}}{1.6(25)\times10^{-7}} & \rates{8.5(18)\times10^{-7}}{4.3(67)\times10^{-10}} & \rates{0}{0} & \rates{0}{0} \\
npop2-cpop05-fb95-nTF-NF & \rates{1.50(30)\times10^{-2}}{5.1(26)\times10^{-3}} & \rates{4.12(82)\times10^{-5}}{1.41(70)\times10^{-5}} & \rates{3.15(68)\times10^{-4}}{4.01(51)\times10^{-6}} & \rates{8.6(19)\times10^{-7}}{1.10(14)\times10^{-8}} & \rates{0}{0} & \rates{0}{0} \\
npop2-cpop2-fb10-TF-NF & \rates{9.88(53)\times10^{-3}}{2.311(84)\times10^{-3}} & \rates{2.71(15)\times10^{-5}}{6.33(23)\times10^{-6}} & \rates{1.666(37)\times10^{-3}}{3.89(17)\times10^{-4}} & \rates{4.56(10)\times10^{-6}}{1.067(45)\times10^{-6}} & \rates{0}{0} & \rates{0}{0} \\
npop2-cpop2-fb10-nTF & \rates{1.161(83)\times10^{-2}}{1.33(14)\times10^{-3}} & \rates{3.18(23)\times10^{-5}}{3.65(39)\times10^{-6}} & \rates{1.24(95)\times10^{-4}}{1.0(14)\times10^{-5}} & \rates{3.4(26)\times10^{-7}}{2.9(37)\times10^{-8}} & \rates{0}{0} & \rates{0}{0} \\
npop2-cpop2-fb10-nTF-NF & \rates{2.9(12)\times10^{-3}}{8.7(24)\times10^{-5}} & \rates{7.9(34)\times10^{-6}}{2.39(67)\times10^{-7}} & \rates{2.2(19)\times10^{-6}}{1.6(21)\times10^{-6}} & \rates{5.9(52)\times10^{-9}}{4.5(57)\times10^{-9}} & \rates{0}{0} & \rates{0}{0} \\
npop2-cpop2-fb95-TF-NF & \rates{7.57(36)\times10^{-2}}{2.65(47)\times10^{-2}} & \rates{2.23(13)\times10^{-4}}{8.8(20)\times10^{-5}} & \rates{4.26(13)\times10^{-3}}{4.10(17)\times10^{-4}} & \rates{1.168(36)\times10^{-5}}{1.124(46)\times10^{-6}} & \rates{1.4(28)\times10^{-7}}{0} & \rates{3.8(76)\times10^{-10}}{0} \\
npop2-cpop2-fb95-nTF & \rates{3.07(45)\times10^{-2}}{7.7(47)\times10^{-3}} & \rates{8.6(15)\times10^{-5}}{2.3(15)\times10^{-5}} & \rates{4.6(10)\times10^{-4}}{3.7(26)\times10^{-6}} & \rates{1.26(28)\times10^{-6}}{1.02(70)\times10^{-8}} & \rates{0}{0} & \rates{0}{0} \\
npop2-cpop2-fb95-nTF-NF & \rates{7.4(16)\times10^{-3}}{7.5(21)\times10^{-4}} & \rates{2.01(44)\times10^{-5}}{2.05(59)\times10^{-6}} & \rates{3.7(30)\times10^{-5}}{5.6(45)\times10^{-6}} & \rates{1.01(83)\times10^{-7}}{1.5(12)\times10^{-8}} & \rates{0}{0} & \rates{0}{0} \\
\bottomrule
\end{tabular}

\end{sidewaystable*}


\end{document}